\documentclass[a4]{article}

\usepackage{amsmath}
\usepackage{MnSymbol}%
\usepackage{wasysym}%
\usepackage{cite}
\usepackage{graphicx}
\usepackage{subfigure}
\usepackage{supertabular}
\usepackage{enumitem}
\usepackage{hyperref}

\usepackage{color}
\usepackage[normalem]{ulem}

\newcommand{\R}[1]{{\color{black}#1}}

\newcommand{\Fref}[1]{Figure \ref{#1}}
\newcommand{\Sref}[1]{Section \ref{#1}}
\newcommand{\Tref}[1]{Table \ref{#1}}
\newcommand{\Eref}[1]{Table \ref{#1}}

\newcommand{\dvol}{{\rm d}^3{\bf r}}
\newcommand{\dsur}{{\rm d}{\bf s}}
\newcommand{\dsec}{{\rm d}^2{\bf r}}

\newcommand{\vJ}{{\bf J}}

\newcommand{\vE}{{\bf E}}
\newcommand{\vB}{{\bf B}}
\newcommand{\vA}{{\bf A}}

\newcommand{\ve}{{\bf e}}

\newcommand{\vr}{{\bf r}}

\begin{document}
\title{Coupling loss at the end connections of REBCO stacks: 2D modelling and measurement}

\author{Shuo Li$^{1,2}$, Enric Pardo$^2$\R{\footnote{\R{Author to whom correspondence should be sent. Email: enric.pardo@savba.sk}}},  J{\'a}n Kov{\'a}{\v c}$^2$ \\
$^1$ College of Information Science and Engineering,\\
Northeastern University, 110004 Shenyang, China.\\
$^2$ Institute of Electrical Engineering, Slovak academy of sciences,\\
84104 Bratislava, Slovakia.
}

\vspace{10pt}

\maketitle

\begin{abstract}
	
In high power density superconducting motors, superconducting tapes are usually stacked and connected together at terminals to improve the current capacity. When a parallel sinusoidal magnetic field \R{is} applied on this partially coupled stack, the coupling current is induced and causes additional coupling loss. Usually 3D \R{modeling} is needed to calculate the coupling loss but it takes too much computing resource and time. In this paper, a numerical 2D modeling by minimum electromagnetic entropy production (MEMEP) method is developed to speed up the calculation. The presented MEMEP model shows good accuracy and the capability to take the realistic resistance between tapes into account for coupling loss calculation with a high number of mesh element, which \R{agrees to measurements}.Thanks to the model, a systemic study of coupling loss on amplitude-dependence, frequency-dependence, resistance-dependence, and length-dependence, is presented and discussed.
The results reveal the features of coupling loss which is very helpful \R{devices with multi-tape conductors, such as the stator or rotor windings of motors}.
\end{abstract}


\vspace{1pc}
%
%
%


\section{Introduction}

Since the discovery of high temperature superconductivity over 30 years ago, there has been great interest in utilizing high temperature superconducting materials to develop high power density motor for aircraft to reduce the pollutant emission of airplanes \cite{haran17SST,ASuMED,pardo2019ac}. \R{In the stator of superconducting motors, using closely packed parallel tapes as conductor improves the current capacity while keeping high engineering current densities \cite{pardo2019ac}. Isolating the tapes along their length reduces the AC loss compared to non-insulated tapes, also when the tapes are connected at the current terminals \cite{grinenko12SST,pardo2019ac}.} This \R{we name this} kind of stack is \R{as} partially coupled stack.

\begin{figure}
\centering		
\subfigure[Uncoupled stack]{
		\centering
		\includegraphics[scale= 0.2]{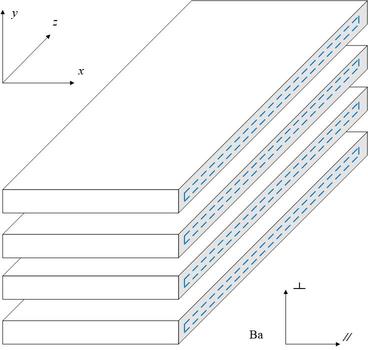}	
		\label{F:stack:a}   }
\quad	
\subfigure[Partially coupled stack]{
		\centering
		\includegraphics[scale= 0.2]{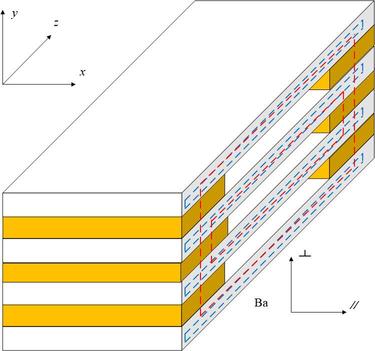}
		\label{F:stack:b}   }
\quad	
\subfigure[Fully coupled stack]{
		\centering
		\includegraphics[scale= 0.2]{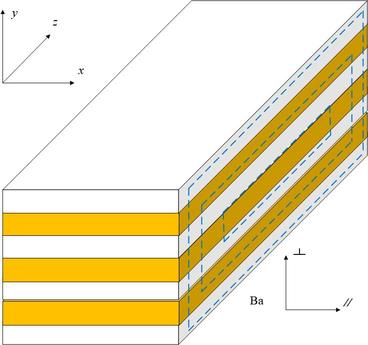}	
		\label{F:stack:c}   }
\caption	{
The sketch of proposed three kinds of superconducting stack.
$(a)$ Uncoupled stack, inducting current flows in tapes independently. 
$(b)$ Partially coupled stack, inducting current flows in tapes parallel but coupled at terminals.
$(c)$ Fully coupled stack, inducting current flows in the entire cross section.
}
\label{F:stack}
\end{figure}

A sketch of the partially coupled stack is shown in \Fref{F:stack:b}. When a parallel \R{time-varying} magnetic field \R{is} applied to the partially coupled stack, coupling current is induced and additional coupling loss is generated because of \r{these} connections. \R{Often,} the coupling loss is thousands times higher than superconducting loss. To predict the coupling loss, usually a 3D model is necessary to solve this problem \cite{couplingEUCAS,mifune19SST}. Right now, \R{commercial finite-element methods (FEM) with} H-formulation and T-A formulation are very popular and flexible in superconducting modeling \cite{brambilla07SST,zhangH17SST,liu2018comparison,berrospe2019real,wang2018study}. However, it takes too much computing resource and time, which is not suitable to model the coupling loss of the partially coupled stack with realistic resistance at terminals.  

In this study, a numerical 2D model by Minimum Electro-Magnetic Entropy Production method (MEMEP) with consideration of realistic resistance between tapes is developed to solve the 3D problem. The presented MEMEP model shows good accuracy and capability, which is verified by measurements. \R{The} MEMEP method is much faster than the H-formulation and T-A formulation methods, so that it can adopt much better mesh strategy to get high precision results \cite{pardo2015electromagnetic, pardo20153d, kapolka2018three,memep3D,pardo2007current, pardo2013calculation}. \R{Although MEMEP has been applied to full 3D problems, it requires either high computing resources (several computers working in parallel by MPI) \cite{memep3D,anisotropy3d}) or high computing time. Therefore, fast and accurate 2D modeling enables analyzing complex structures, such as stacks of many tapes or coils made of these stacks.
}

Besides, when the resistance between tapes is extremely large, the tapes are fully insulated from each other, which is called uncoupled stack, as seen \R{in} \Fref{F:stack:a}. Conversely, when the resistance is  extremely small, the tapes are connected very well from each other, which is called fully coupled stack, as seen \R{in} \Fref{F:stack:c}. \R{Althgough}, uncoupled stacks \R{(and coils)} and fully coupled stacks \R{have been} well studied by many researchers before\R{, the case of an intermediate finite resistance (``partial coupling") has not been previously studied for stacks of tapes}. \R{Previous studes focused} on 
ac loss under perpendicular magnetic field \R{\cite{pardo03PRB, grilli2006magnetization, kajikawa09PhC, hongZ10JSM, pardo12SSTa, bykovsky2017magnetization, ueno2017ac, climente2019computation, yazdani2018investigation}}, ac loss under transport current \R{\cite{pancaketheo,prigozhin11SST,ainslie2012calculating,zermeno2013calculation, song2018numerical}}, ac loss with consideration of ferromagnetic material \R{\cite{eucas10fmsc,pancakeFM,liu2017numerical, ogawa2016experimental}}, dynamic resistance under various applied field \R{\cite{tapeDCAC,jiang2017dynamic,ainslie18SST,liu2019dynamic}}, magnetization and decay of trapped magnetic field \R{\cite{ainslie2015modelling, zouS17SST, campbell17SST, liangF17SST, baghdadi18ScR, li2019decay, cubedemagnetization, stackdemagnetization3d}}, multifilamentary calculations \R{\cite{pardo03PRB,amemiya2004ac}}, \R{among other topics}.

Thanks to the presented numerical 2D MEMEP model, \R{this article systematically studies and discusses} the amplitude-dependence, frequency-dependence, and resistance-dependence of coupling loss under parallel magnetic field. The results reveal the features of coupling loss\R{,} which \R{will be} very helpful in \R{future design considerations of power applications, such as superconducting motors, and magnets}. \R{However, rather than the particular modeling results, the main innovation of this article is the fast and accurate 2D modeling method and developed software to take coupling currents into account, avoiding complex time-consuming 3D modeling.}

The \R{structure} of this paper is \R{the following}. The details of superconducting samples are described and measurement setup is presented in section \ref{sec:sample}.
Then, the general rule of 2D minimum electromagnetic entropy production method and coupling loss modeling for the partially stack with realistic resistance at terminals are presented in  \Sref{sec:MEMEP} and \Sref{sec:coupling}, respectively. 
Magnetization loss of uncoupled and partially coupled stack samples are measured under applied sinusoidal magnetic field with a series of amplitude and frequency to verify the presented numerical 2D MEMEP model. 
Then, the amplitude-dependence, frequency-dependence, and resistance-dependence of coupling loss of partially coupled stack under parallel magnetic field are systematically discussed in \Sref{sec:Result}.  
At last, some conclusions are given in \Sref{sec:conclusion}.
In the appendix, an analytical modeling of coupling loss for partially stack with four tapes at low frequency and with two tapes for any frequency are deduced in \ref {sec:app_low} and \ref{sec:app_high}, respectively.


\section{Samples and measurement setup} \label{sec:sample}
\subsection{Samples}

For this work, \R{REBCO} superconducting stacks are systematically investigated (\R{REBCO stands for $RE$Ba$_3$Cu$_3$O$_{7+x}$, where $RE$ is} a rare earth element, \R{usually yttrium or gadolinium}). 
The nonmagnetic superconducting tapes are provided by SuperOx, \R{whose} critical current is about $160$ A at $77$ K under self-field, and the cross section is $4$ mm width \R{times} $0.2$ mm thick.

\begin{figure}	
	\centering
	\includegraphics[scale= 0.7]{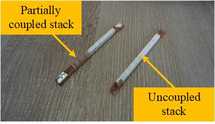}   
	\caption	{
A photograph of the superconducting stack samples.
}	
	\label{F:sample}
\end{figure}

Two $100$ mm long superconducting stack samples are fabricated. One sample is \R{an} uncoupled stack, \R{where} four superconducting tapes are \R{vertically} ``Face to Back" stacked together with $1.5$ mm thick G10 to keep the gap between tapes. Another sample is \R{a} partially coupled stack with the same configuration as uncoupled stack, but the superconducting tapes are soldered together with conventional conductor at terminals with measured average resistance $R_{av}=3.13 \times 10^{-6}$ $ \Omega$. 
The soldering parts length is about $5$ mm on each side of terminals. The detail parameters of superconducting tape and stacks are given in \Tref{tab:sample} and a photograph of stack samples is shown in \Fref{F:sample}

\begin{table} 
\centering
\caption{ \label{tab:sample} The specific parameters of superconducting tape and stacks}		
\begin{tabular}{@{}lr}
	\hline
  \hline		
Parameters                                &  Value \\
  \hline		
Critical current [A]              &    160 \\
$n$-value                                 &     30 \\
Width of tape [mm]                        &      4 \\
Thickness of tape [mm]                     &    0.2 \\
Thickness of superconductor layer [$\mu$m] &      2 \\
Number of tapes in stack                  &      4 \\
Gap of tapes [mm]                         &    1.5 \\
Length of stacks [mm]                     &    100 \\
Average resistance at ends $[\mu \Omega]$ & $3.13$ \\
  \hline                                    &
\end{tabular}	    
\end{table}


\subsection{Measurement setup}

\begin{figure}	
	\centering
	\includegraphics[scale= 0.6]{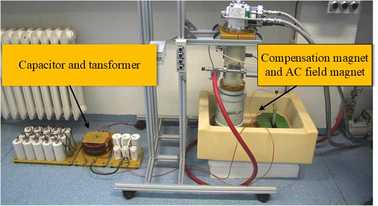}   
	\caption	{
		A photograph of the measurement system}	
	\label{F:measure}
\end{figure}

A photograph of the measurement system is shown in \Fref{F:measure}. A DSP lock-in amplifier 7265 (EG\&G Instruments) is used as a basic measuring device. Then, the signal from its internal generator is utilized as the input signal for power audio amplifier QSC RMX 1450. \R{Copper-wound electro-magnets} are then powered through a toroidal power transformer with ferromagnetic core. The whole coil system is immersed in liquid nitrogen to keep this magnetic system operating at $77$ K. In this way, the change of impedance of the secondary due to its heating is excluded. A sinusoidal external field in the range of $1  \sim 100$ mT and frequency in the range of $2 \sim 576$ Hz is applied by \R{the} background \R{electro-magnet}. The magnetic field homogeneity in the sample space is less than $1\%$. The magnetization loss could be measured directly when the sample is placed in the center of the coil. More detail about this measurement system can be found in  previously work \cite{vsouc2005calibration}.


\section{Minimum Electro-Magnetic Entropy Production method} \label{sec:MEMEP}

Generally, for any \R{non-linear $\vE(\vJ)$ relation of the material and Coulomb's gauge ($\nabla\cdot\vA=0$) for} the vector potential, the current density follows,	
\begin{eqnarray}  \label{E_fai}
\mathbf{E(J)}         &=  -\frac{\partial  \mathbf{ A}_J}{\partial t} - 
\frac{\partial   \mathbf{A}_a}{\partial t}-\nabla \phi.
\end{eqnarray}	  		
where $\mathbf {A}_J$ is the vector potential created by \R{the} current density $\mathbf{J}$\R{. This quantity is \cite{acreview}}
\R{
\begin{equation}\label{A-J3D}
\mathbf { A}_J \mathbf{ (r) }
= \frac{\mu_0}{4\pi} \int_{\Omega_{3D}}  \frac{ \mathbf {J(r')}}{|  \mathbf {r} -   \mathbf {r'}|}  \dvol ' 
\end{equation} 
for three-dimensional (3D) modeling, where $\mu_0$ is the vacuum permeability\footnote{In this article, we take a value of the vacuum permeability of $\mu_0=4\pi\cdot 10^{-7}$ H/m. Strictly speaking, this is not an exact value because, according to the 2019 definition of SI, $\mu_0$ is a measured constant.}, $\vr$ and $\vr'$ are the position of the observation point and generating point, respectively, and $\Omega_{3D}$ is the region where there are currents of the 3D object of study. 
}

\R{The second item in \Eref{E_fai}, $\mathbf{A}_a$, is the ``applied vector potential" or vector potential created by any applied magnetic field, $\mathbf{B}_a$, due to sources (currents or magnetic poles) that are external to the sample domain $\Omega_{3D}$ and independent on the currents in $\Omega_{3D}$. In general, $\vB_a$ does not need to be uniform. In case of uniform $\vB_a$, with components only in the $x$ and $y$ directions, the applied vector potential can be written as
\begin{equation} 
\vA_a=(B_{a,x}y-B_{a,y}x){\bf e}_z,
\label{E:Aa}
\end{equation}
which follows Coulomb's gauge. Although there are other possible ways to express this vector potential, such as $\vA_a=\vB_a\times\vr$, we choose the option above for convenience later one.
}

Solving the equation above is the same as minimizing the following functional for the change of current density \R{between two time steps} \R{\cite{memep3D}}. \R{After time discretization, this functional is}
\R{
\begin{eqnarray} \label{E:L}
F = &\int_{\Omega_{3D}} \Bigg[ \frac{1}{2} \frac{\Delta \mathbf{A}_J}{\Delta t} \cdot \Delta \mathbf{J}+ 
\frac{\Delta \mathbf{A}_a}{\Delta t}\cdot \Delta \mathbf{J}              \nonumber \\
& + U(\mathbf{J}_0+\Delta \mathbf{J})  + \nabla \varphi \cdot (\mathbf{J}_0+\Delta \mathbf{J}) \Bigg] \mathrm{\dvol}
\end{eqnarray}
}
where, $\vJ_0$ and $\vJ$ are the current density at times $t=t_0$ and $t=t_0+\Delta t$, respectively \R{(This $\Delta t$ does not need to be uniform)}, $\Delta \mathbf{J}= \mathbf{J}- \mathbf{J}_0$ is \R{the change of} current density \R{between} two time steps\R{, $\Delta \vA_J / \Delta t$} is the average time derivative between $t=t_0$ and $t=t_0+\Delta t$ of the vector potential generated by the current density in the sample, $\Delta \mathbf{A}_a / \Delta t $ is the same quantity but relative to the applied vector potential\R{, and $\varphi$ is the electrostatic scalar potential}.	

The dissipation factor \R{in (\ref{E:L}),} $U(\mathbf J)$, is defined as,
\begin{eqnarray} \label{E:U}
U(\mathbf {J}) \equiv \int_{0}^{\vJ} \mathrm{d} \mathbf {J}' \cdot  \mathbf {E (J')}
\end{eqnarray}
which is uniquely defined because \R{$\nabla_{\vJ} \times \mathbf {E (J)}=0$} for any physical $\mathbf {E (J)}$\R{, and hence $U$ follows the property that $\nabla_{\vJ}U(\vJ)=\vE(\vJ)$} \cite{pardo2015electromagnetic}.
For a small $\Delta \mathbf{J}$, the dissipation factor is a measure of the energy dissipation due to $\Delta \mathbf {J}$, 
since $U (\mathbf{J}_0 + \Delta \mathbf {J}) - U(\mathbf {J}_0) \approx \Delta \mathbf {J} \cdot \mathbf {E(J}_0  \mathbf )$. \R{With this definition of the dissipation factor, it can be seen that the functional of \Eref{E:UJ} always presents a minimum and it is unique \cite{memep3D}.}

Usually, superconductors can be approximated by a Power Law $  \mathbf {E (J)}$ relation \R{and} the \R{conventional} conductor could be approximated by a linear $ \mathbf {E (J)}$ relation\R{; and hence} 
\begin{eqnarray} \label{E:PL}
\mathbf {E (J)} =
\begin{cases} 
\frac{E_c}{J_c}  {\left| \frac{J}{J_c}\right|}  ^{n-1} \mathbf{J}, & \text{for the superconductor} \\
\rho \cdot \mathbf {J},   &\text{for the conventional conductor}
\end{cases}
.
\end{eqnarray}
\R{Above, $E_c$ defines the critical current criterion, since $|\vE|=E_c$ when $|\vJ|=J_c$, being $E_c$ usually set as $E_c = 10^{-4}$ V/m;} $J_c$ is critical current density of superconductors\R{;} and $\rho$ is a constant conductivity. Thus, the dissipation factor \R{of} \Eref{E:PL} becomes,
\begin{eqnarray} \label{E:UJ}
U \left( \mathbf{J} \right)=
\begin{cases}
E_c \frac{J_c}{n+1}  \left|\frac{\mathbf{J}}{J_c}\right|^{n+1}  & \text{for the superconductor}\\
\frac{1}{2}  \rho \mathbf{J}^2 & \text{for the conventional conductor} \R{.} 
\end{cases}
\end{eqnarray}

\R{
For shapes infinitely long in the $z$ direction, the current density follows $\vJ(\vr)=J(x,y)\ve_z$, and hence $\vA(\vr)_J=A_J(x,y)\ve_z$ and $\vE(\vr)=E(x,y)\ve_z$. From this, $\vA_a$ of \Eref{E:Aa}, and \Eref{E_fai} we also see that $\nabla\varphi(\vr)=\partial_z\varphi\ve_z$ with uniform $\partial_z\varphi$ for each conductor (for multiple conductors). If the conductors are all of the same length $l$, $\partial_z\varphi$ is related to the voltage drop along the whole length, $V$, as $V=-l\partial_z\varphi$, where we define the positive current sign as that following the positive $z$ direction. Then, the functional of \Eref{E:L} becomes
\begin{equation} \label{E:F2D}
F=l\int_{\Omega_{2D}} \Bigg[\frac{1}{2}\frac{\Delta A_J}{\Delta t}\Delta J + \frac{\Delta A_a}{\Delta t}\Delta J + U(J_0+\Delta J)\Bigg]\dsec - \sum_i^{n_t} V_i I_i .
\end{equation}
Above, $\Omega_{2D}$ is the domain comprising the cross-section of all conductors $n_t$ is the number of conductors (or tapes, in our case); $i$ labels each conductor; and $V_i$ and $I_i$ are the voltage drop and net current for each conductor, respectively, at time $t_0+\Delta t$, being the time that we solve the current density. In \Eref{E:F2D}, $A_J$ in Coulomb's gauge can be written as \cite{acreview}
\begin{equation}
A_J(\vr)=-\frac{\mu_0}{2\pi}\int_{\Omega_{2D}}\dsec J(\vr)\ln|\vr-\vr'|.
\end{equation}
If the current constrains, $I_i$, are given and the minimization method respects the current constrain, the last term of \Eref{E:F2D} could be dropped from the functional \cite{pardo2015electromagnetic}. However, for the partially coupled case, these constrains are not fixed, and hence we needed to develop the method explained in the next section.
}


\section{Coupling loss modeling} \label{sec:coupling}

The magnetization loss of \R{the two coupling limits of} uncoupled stack and fully coupled stack can be computed by \R{minimizing the functional of} \Eref{E:F2D} directly with some additional \R{conditions}  to distinguish these two stacks \R{\cite{pardo2015electromagnetic}}. \R{However, we need to take further considerations when the tapes are joined by a finite resistance (partial coupling), as follows.}

For partially coupled stack, superconducting tapes are connected together at terminals and the coupling current is induced when a parallel \R{time-varying} magnetic field is applied.
The induced current appears not only in \R{the} superconductors but also in the resistance between tapes.
The coupling current causes coupling loss \R{and modifies the net current in each superconductor}. To \R{take the coupling currents into account, we should consider the full 3D formulation of \Eref{E:L}. However, we can make several important simplifications, as follows. First, we can separate the current-density into that from the superconductor, $J_s$, and that in the resistive material, $J_R$. When doing so, the functional of \Eref{E:L} can be separated into}
\R{
\begin{eqnarray} \label{E:Gamma_c}
F=F_{S}+F_{R}+F_{RS},
\end{eqnarray}}
where $F_{S}$ \R{and $F_R$ are} the \R{functional} contributed from \R{the} superconductor \R{and resistive material, respectively, and $F_{SR}$ represents their mutual interaction.}

\R{Since the superconducting tapes are very long, we can use the 2D infinitely long assumption for $F_S$, resulting in [see \Eref{E:F2D}]
\begin{eqnarray} \label{E:L_S}
F=l_s\int_{\Omega_{2D,s}} \Bigg[\frac{1}{2}\frac{\Delta A_s}{\Delta t}\Delta J_s + \frac{\Delta A_a}{\Delta t}\Delta J_s + U(J_{s0}+\Delta J_s)\Bigg]\dsec - \sum_i^{n_t} V_i I_i .
\end{eqnarray}}
where $l_s$ is the length of \R{each tape, $\Omega_{2D,s}$ is the cross-sectional region of all superconducting material, and $\Delta A_s$ is the vector potential created by $\Delta J_s$}.

\R{For the resistive material, we take the 3D functional by now, being
\begin{eqnarray} \label{E:L_R0}
F_R = &\int_{\Omega_{3D,R}} \Bigg[ \frac{1}{2} \frac{\Delta \mathbf{A}_R}{\Delta t} \cdot \Delta\vJ_R+ 
\frac{\Delta\vA_a}{\Delta t}\cdot \Delta\vJ_R              \nonumber \\
& + U(\vJ_R)  + \nabla \varphi \cdot \vJ_R \Bigg] \mathrm{\dvol},
\end{eqnarray}
where $\Omega_{3D,R}$ is the 3D domain where there is resistive material, $\Delta\vA_R$ is the vector potential generated by $\Delta\vJ_R$, and $\vJ_R$ is the current density in the resistive material at time $t_0+\Delta t$. Next, we assume large enough resistivity (or low enough frequency) to neglect dynamic effects. Then, 
\begin{eqnarray} \label{E:FRst}
F_R\approx \int_{\Omega_{3D,R}}\dvol U(\vJ_R) + \int_{\Omega_{3D,R}}\dvol \nabla\varphi\cdot\vJ_R.
\end{eqnarray}
Next, we consider a single resistance only. From \Eref{E:PL}, the dissipation factor follows $U(\vJ_R)=\rho\vJ_R^2/2$, which due to Ohms law ($\vE=\rho\vJ$) it beocomes $U(\vJ_R)=\vE\cdot\vJ_R/2$. Since this is the local dissipation, the volume integral of $\vE\cdot\vJ_R$ needs to equal to the power dissipation of the resistance, being $RI_R^2$. Then, the first integral in \Eref{E:FRst} is $RI_R^2/2$ for any shape of the resistive connection. The second term in \Eref{E:FRst} can also be simplified. Using $\nabla\cdot(\varphi\vJ_R)=\nabla\varphi\cdot\vJ_R+(\nabla\cdot\vJ_R)\varphi$ and $\nabla\cdot\vJ_R=0$, the second integral of \Eref{E:FRst} becomes
\begin{eqnarray} \label{E:VRI}
\int_{\partial\Omega_{3D,R,1}}\varphi\vJ_R\cdot\dsur = (\varphi_o-\varphi_i)I_R=-V_RI_R,
\end{eqnarray}
where $\Omega_{3D,R,1}$ is the region where a single resistance exists. At the first equality of \Eref{E:VRI} we used that the input and output surfaces of the current are equipotentials and at the second equality we used the standard definition of the voltage drop in the resistance, $V_R$. Therefore, under the static assumption \Eref{E:L_R0} for a single resistance becomes
\begin{eqnarray} \label{E:FR}
F_{R}\approx \frac{1}{2}RI_R^2 - V_RI_R
\end{eqnarray}
for any 3D shape of the resistive material with overall resistance $R$. If there are $n_R$ resistances, the total functional for the resistances is 
\begin{eqnarray} \label{E:FR}
F_{R}\approx \sum_j^{n_R} \Bigg[ \frac{1}{2}R_jI_j^2 - V_jI_j \Bigg].
\end{eqnarray}
}

\begin{figure}
	\centering	
	\subfigure[$x-y$ plane]{
		\centering
		\includegraphics[scale= 0.35]{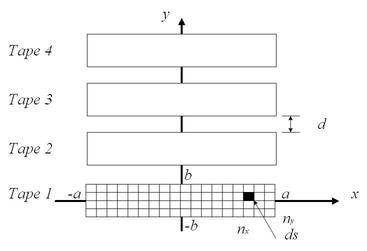} 
		\label{F:Stack:xy} }
	\quad
	\subfigure[$y-z$ plane]{
		\centering
		\includegraphics[scale= 0.354]{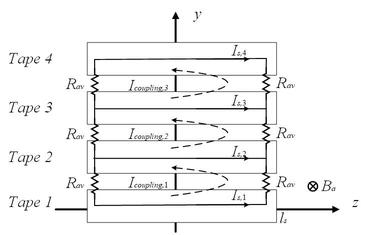} 
		\label{F:Stack:yz} }
	\caption	{
		Sketch of modeling of partially coupled stack with resistance	
	}
	\label{F:Stack}
\end{figure}

\R{The interaction term, $F_{RS}$, is
\begin{eqnarray}\label{E:FRS}
F_{RS} & = & \int_{\Omega_{3D}}\dvol\Bigg[ \frac{1}{2} \Delta\vJ_s\cdot\frac{\Delta\vA_R}{\Delta t} + \frac{1}{2} \Delta\vJ_R\cdot \frac{\Delta\vA_s}{\Delta t} \Bigg] \nonumber \\
& = & \int_{\Omega_{3D}}\dvol \Delta\vJ_s\cdot\frac{\Delta\vA_R}{\Delta t} \approx 0.
\end{eqnarray}
The last integral can be neglected, since $\vJ_s$ in the superconductor (and the $\vA_s$ that it generates) mostly follows the $z$ direction, while $\vJ_R$ (and its generated $\vA_R$) is mostly perpendicular to the $z$ axis; and hence the dot product in the last integral above cancels.
}

\R{Therefore, the total functional for very long tapes and neglecting dynamic effects in the resistances is
\begin{eqnarray}\label{E:Fcoubare}
F & \approx & \int_{\Omega_{2D,s}}\Bigg[\frac{1}{2}\frac{\Delta A_s}{\Delta t}\Delta J_s + \frac{\Delta A_a}{\Delta t}\Delta J_s + U(J_{s0}+\Delta J_s)\Bigg]\dsec \nonumber \\
&& + \sum_{j=1}^{n_R}\frac{1}{2}R_jI_j^2 - \sum_{i=1}^{n_t} V_iI_i - \sum_{j=1}^{n_R}V_jI_j.
\end{eqnarray}
Since the resistive material is at the ends of the tapes, the current in each superconductor or resistance forms closed loops. Figure \ref{F:Stack} shows our assumed circuits for the studied stacks, where only neighboring tapes are directly connected, but the following reasoning applies for any cross-connection between tapes. For any cross-connection, the last two terms in \Eref{E:Fcoubare} can be grouped in closed loops, and hence they vanish due to the voltage Kirshoff law. For the example of figure \ref{F:Stack}, these two terms become
\begin{eqnarray}\label{E:loop}
- \sum_{k=1}^{n_l} [ V_{sk}-V_{s,k+1}+V_{Rk+}-V_{Rk-} ] I_{{\rm coupling},k},
\end{eqnarray}
where $I_{{\rm coupling},k}$ is the coupling loop current of a particular loop $k$, and $V_{Rk+}$, $V_{Rk-}$ are the voltages at the right and left resistances of loop $k$, respectively. When writing the sum of voltages above as differences of potentials, it can be easily seen that it vanishes. Then, the functional to minimize is reduced to
\begin{eqnarray}\label{E:Fcou}
F & \approx & \int_{\Omega_{2D,s}}\Bigg[\frac{1}{2}\frac{\Delta A_s}{\Delta t}\Delta J_s + \frac{\Delta A_a}{\Delta t}\Delta J_s + U(J_{s0}+\Delta J_s)\Bigg]\dsec \nonumber \\
&& + \sum_{j=1}^{n_R}\frac{1}{2}R_jI_j^2.
\end{eqnarray}
For our present case of figure \ref{F:Stack}, the last sum can be easily written in terms of loop currents, with a resulting functional
\begin{eqnarray}\label{E:Fstack}
F & \approx & \int_{\Omega_{2D,s}}\Bigg[\frac{1}{2}\frac{\Delta A_s}{\Delta t}\Delta J_s + \frac{\Delta A_a}{\Delta t}\Delta J_s + U(J_{s0}+\Delta J_s)\Bigg]\dsec \nonumber \\
&& + \sum_{j=1}^{n_t-1}R_{av}I_{{\rm coupling},j}^2.
\end{eqnarray}
The loop currents are related to the currents at each superconducting tape, $I_i$ as
\begin{eqnarray}\label{E:Icou}
I_{{\rm coupling},k}=
\begin{cases}
-I_1  & \mbox{for $k=1$}\\
I_{k-1}-I_k & \mbox{for $1<k<n_t-1$} \\
I_{n_t} & \mbox{for $k=n_t-1$}
\end{cases}
.
\end{eqnarray}
Therefore, the coupling loop currents are not independent variables. The reader should note that any coupling current distribution between tapes can be expressed as a linear combination of our defined loop coupling currents. For instance a loop flowing between the first and last tape corresponds to the case when the loop current in all unit loops is the same.
}

\R{In order to find the solution of the current density by minimizing \Eref{E:Fstack}, we divide the tape cross-section into elements, being $N$ the total number of elements in all tapes.} To minimize \Eref{E:Fstack}, we can use \R{an algorithm similar to that in \cite{pardo2015electromagnetic}. Outlining, the algorithm follows the main steps below:
\begin{enumerate}[label={\arabic*.}]

\item For each tape $p$, find the element pair within that tape, $i_{p+}$ and $i_{p-}$, where applying a change in current $+\delta I$ and $-\delta I$ (with positive $\delta I$), respectively, causes the smallest increment in $F$, $\delta F_j$. For this, neglect the interaction between $i_{p+}$ and $i_{p-}$, which can always be done for a high enough number of elements.

\item For each coupling loop, $k$, consisting of its upper and lower tape and its right and left resistances, find the element pair, $i_{k+}$ and $i_{k-}$, with $i_{k+}$ and $i_{k-}$ belonging to different tapes where applying a change in current $+\delta I$ and $-\delta I$, respectively, causes the smallest increment in $F$, $\delta F_k$. Again, we neglect the interaction between $i_{k+}$ and $i_{k-}$ but now we need to take the resistance term in \Eref{E:Fstack} into account. Naturally, for open circuit, where $R_{av}$ is infinite, $\delta F_k$ is also infinite, which will prevent setting currents in the coupling loops.

\item Choose the element pair, either within the same superconductor or in a coupling loop, where applying the change of current minimizes the functional the most. 

\item If the change in the functional is negative, set this change of current in the related pair of elements. Then, calculate the new current in all tapes; and in all loops, if required. Any algorithm accelerators, such as the vector potential in the elements, should be calculated here, as done in \cite{pardo2015electromagnetic}. Afterwards, repeat the process from step 1.

\item If the change of functional is non-negative, it means that we reached the unique minimum. Therefore, the routine ends.

\end{enumerate}
For magnetic field dependent $J_c$ (or other magnetic-field dependent parameters), we can solve the current density by combining this routine with the calculation of the magnetic field in an iterative way, as done in \cite{pardo2015electromagnetic}.
}

\R{Once the current density is calculated}, the power loss at time $t$, $P(t)$, in the entire cross section is computed for each time step\R{, being} 
\begin{eqnarray} \label{E:P_ALL}
P(t)
& = \R{P_{s}(t)+P_{R}(t)}  \nonumber \\
& =l_s \sum_{i}^{N} J_i(t) E_i(t) s_i + 2 \sum_{k}^{n_{tape}-1} R_{av} I_{{\rm coupling},k}^2 (t),
\end{eqnarray}
\R{where the first and second terms are the superconductor and resistor loss, respectively. In this work, we distinguish between the superconductor and resistive loss, as occurring in the superconductor and in the resistive joints, respectively, instead of hysteresis and coupling loss \cite{acreview}. The reason is that the coupling loss is difficult to define, since the coupling currents in the resistive joint create AC loss not only in the join itself but also modify the AC loss in the superconductor, creating a frequency-dependent superconductor loss.}

The loss Q over one cycle is obtained by \R{the} integration \R{of} \Eref{E:P_ALL}.	
\begin{eqnarray} \label{E:Q}
Q  &=\R{Q_{s} +Q_{R}}   \nonumber \\
&= \int_{mT}^{(m+1)T} P_{s}(t)  \mathrm{d}t  + \int_{mT}^{(m+1)T} P_{R}(t)  \mathrm{d}t     
\end{eqnarray}
where $T$ is \R{the} period and $m$ is the number of periods having \R{been} computed, usually taken 0.5 (or more) for the stabilization of \R{the} calculated $Q$. \R{Again, the first and second terms in \Eref{E:Q} are the superconductor and resistor loss per cycle, respectively.}


\section{Results and discussion} \label{sec:Result}

The electromagnetic properties of the stack are solved by the minimum electro-magnetic entropy production method programmed in C++.

A stack with four tapes is modeled.
The first superconducting tape of the stack is located at \R{$-a < x < a$, $-b < y < b$} and meshed by $n_x$ elements along x-axis and $n_y$ elements along y-axis, respectively. 
Other tapes are located along y-axis with gap $d=1.5$ mm between tapes (see \Fref{F:Stack:xy}).  
\R{Therefore, the total number of elements} is $N=n_{tape} \times n_x \times n_y$, here $n_{tape}=4$.
\R{We use the} actual size of \R{the} superconducting layer with $2a=4$ mm width and $2b=2$ $\mu$m thick is used to model superconductors.	Proper mesh $n_x=5120, n_y=1$ and $n_x = 40, n_y = 64$ are chosen to calculate parallel and  perpendicular applied sinusoidal magnetic field situation, respectively. \R{We also use a tape length of 100 mm.}

Furthermore, an analytical modeling of coupling loss for partially stack with four tapes for low frequency is deduced in \ref {sec:app_low}\R{, which is valid for low frequencies}.
The average coupling power and coupling loss formulas are rewritten here,
\begin{eqnarray}	 \label{EA:power}
<P> = \frac{3\omega ^2}{4R_{av}}(l_s \cdot d \cdot  {B}_m)^2 
\end{eqnarray}
\begin{eqnarray}	 \label{EA:Q}
<Q> = \frac{3 \pi \omega}{2R_{av}}(l_s \cdot d \cdot {B}_m)^2 
\end{eqnarray}
where $R_{av}$ is the average resistance, \R{and} $B_m$ is the amplitude of applied parallel magnetic field.

\subsection{Amplitude dependence}
\begin{figure}
\centering	
\subfigure[Uncoupled stack]{
 	\centering
 	\includegraphics[scale= 0.4]{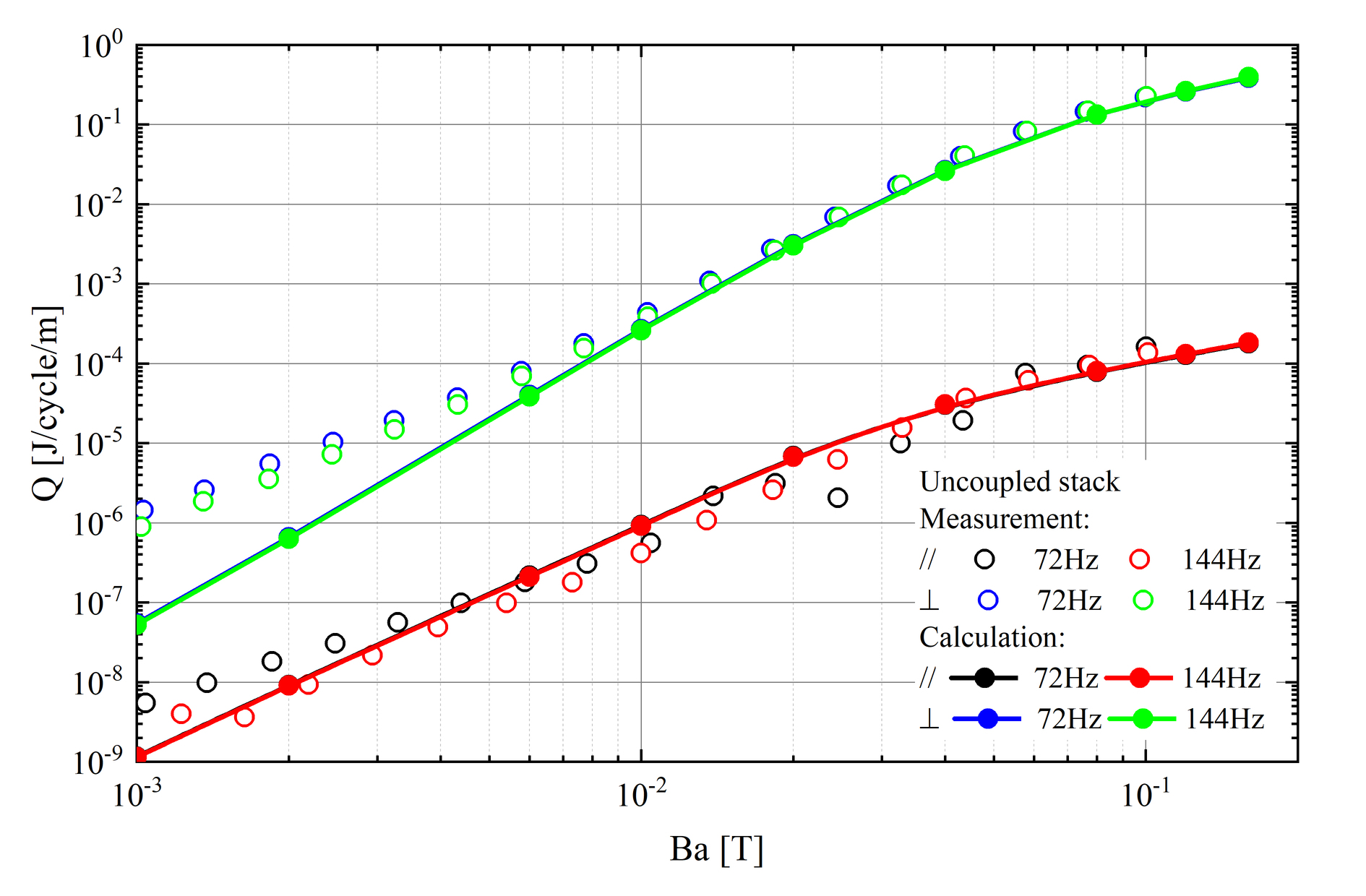} }
   \label{F:amplitude:a}
\quad
\subfigure[Partially coupled stack]{
 	\centering
 	\includegraphics[scale= 0.4]{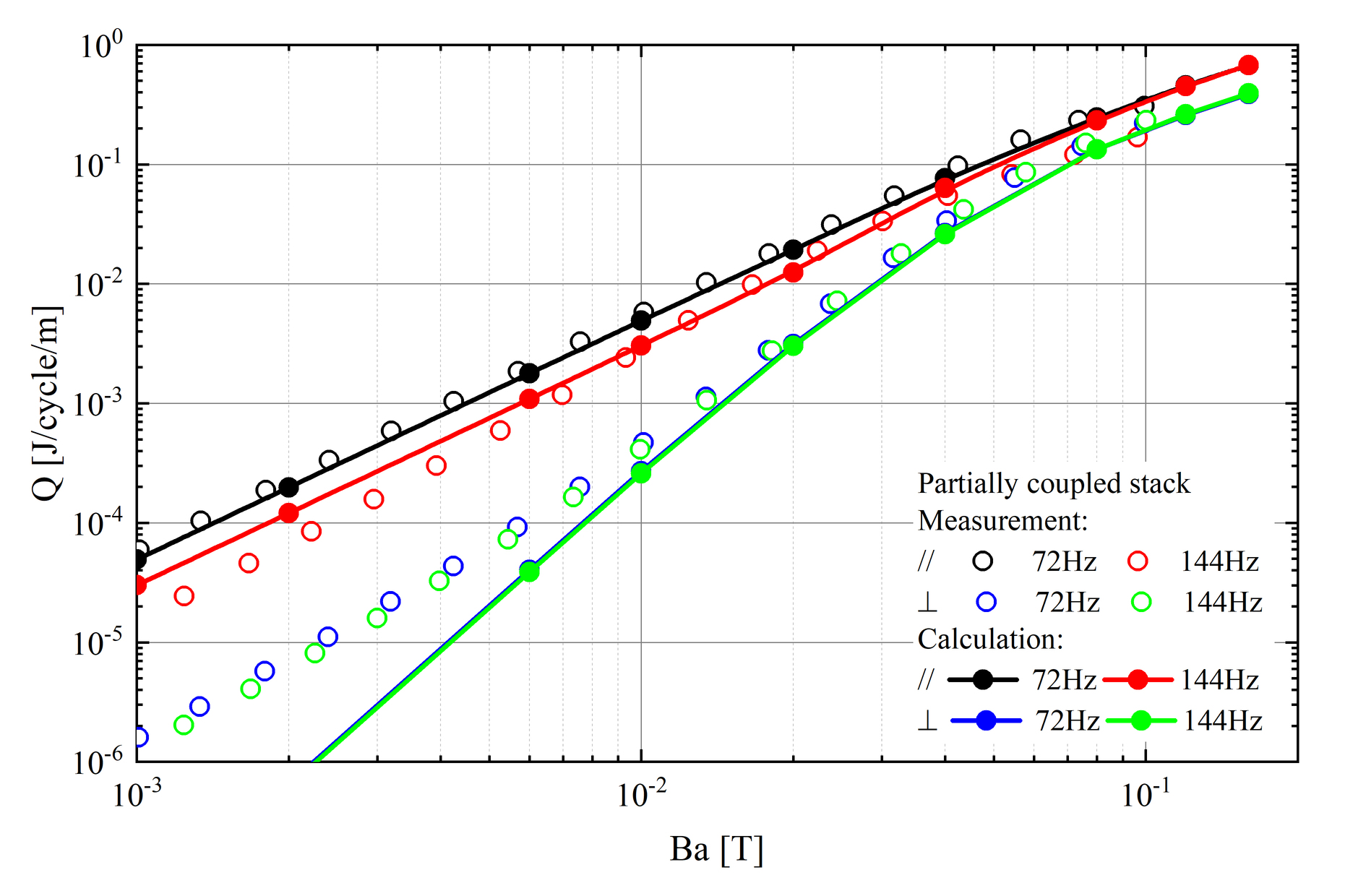}	}
    \label{F:amplitude:b}
 \caption 	{ 
Magnetization loss of uncouple stack and partially coupled stack under parallel and perpendicular magnetic field, respectively.  Loss $Q$ is normalized to J/cycle/m.
The amplitude of applied field increases from $1$ mT to $100$ mT at frequency $72$ Hz and $144$ Hz, respectively. 
The open circles are measurement results and the filled circles are calculation results.
$(a)$ Uncoupled stack.  
$(b)$ Partially coupled stack.		
}
\label{F:amplitude}
\end{figure}

The magnetization \R{loss} of uncoupled stack and partially coupled stack samples are measured and calculated \R{for sinusoidal} magnetic \R{fields of} amplitude ranging from $1$ mT to $100$ mT \R{and} frequency \R{between} $72$ Hz and $144$ Hz. The measurement and calculation results are presented in \Fref{F:amplitude}.

Under parallel magnetic field situation, the induced current flows in $y-z$ plane. For partially coupled \R{stacks}, the induced current flows across the \R{soldered} parts between tapes and causes coupling loss. The normalized magnetization loss in partially coupled \R{stacks} increases from $5 \times 10^{-5} $ to $3 \times 10^{-1} $ J/cycle/m, which is over $3000$ times higher than the loss in uncoupled stack. 
\R{Since} the uncoupled stack and partially coupled stack have the same configuration except the soldering parts, this great difference in magnetization loss mainly contributed by the coupling loss caused by the resistance between tapes.

Under perpendicular magnetic field situation, the current is induced within the superconducting tapes only, which flows in $\pm z$ directions. Thus, partially \R{the} coupled stack has the same current distribution and the same loss as \R{the} uncoupled stack. When the amplitude increases from $1$ mT to $100$ mT, the normalized magnetization loss $Q$ increases from $1 \times 10^{-6} $ to  $2 \times 10^{-1}$  J/cycle/m. At low magnetic field amplitudes, measurements are higher than calculations. This could be caused by a possible degradation of the tape at edges \cite{solovyov2013non}. With the increasing of magnetic field amplitude, the induced current penetrates from edges into the center, and hence edge damage effect is \R{less important than} before. Thanks to this, the calculations are almost the same as the measurements with a relative deviation smaller than $3 \%$.

\begin{figure}	
\centering	
\subfigure[Uncoupled stack]{
		\centering
		\includegraphics[scale= 0.2]{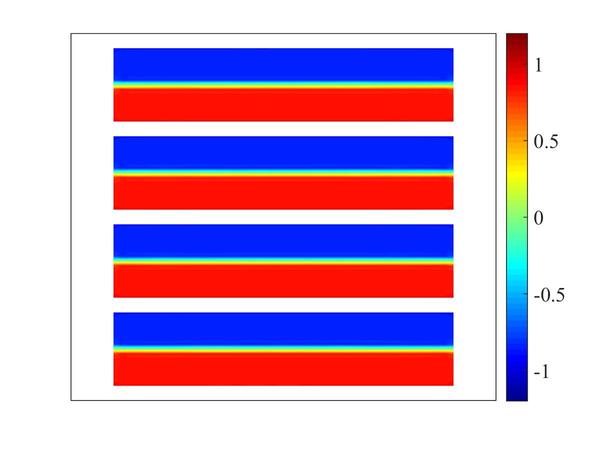}
		\label{F:Cur_2D:a} }
\subfigure[Fully coupled stack]{
		\centering
		\includegraphics[scale= 0.2]{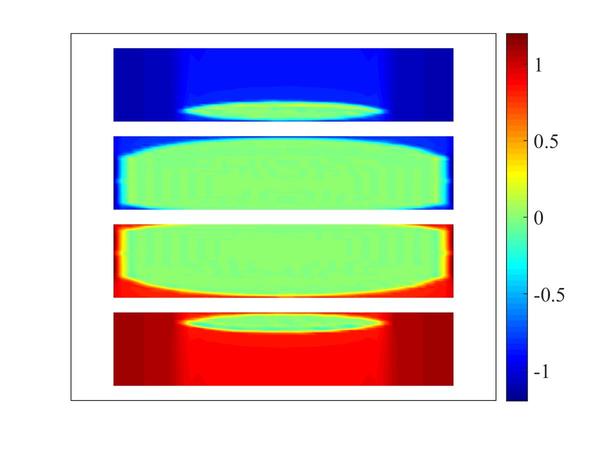} 
		\label{F:Cur_2D:b}}
\subfigure[Partially coupled stack]{
		\centering
		\includegraphics[scale= 0.2]{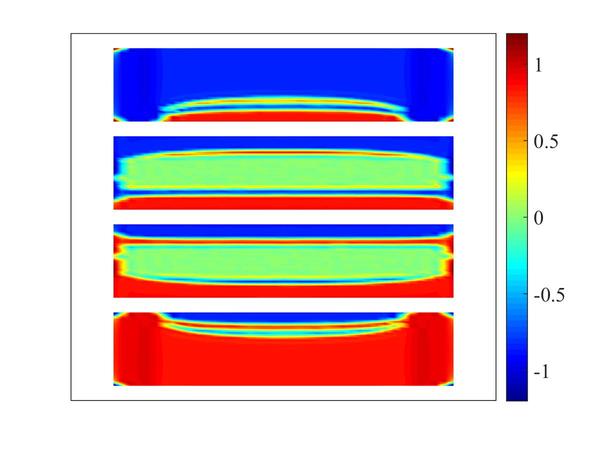} 
		\label{F:Cur_2D:c}}
\caption	{ 
Current distribution of uncouple stack, fully coupled stack and partially coupled stack under parallel magnetic field, respectively when $B_m = 0.02$ T, $f = 72$ Hz, $t/T = 1.25$. 
The thickness of tape is scaled over 100 times to draw the distribution.  
$(a)$  Uncoupled stack. 
$(b)$  Fully coupled stack.
$(c)$  Partially coupled stack.
}
\label{F:Cur_2D}
\end{figure}

\begin{figure}
\centering	
\subfigure[Uncoupled stack]{
		\centering
		\includegraphics[scale= 0.2]{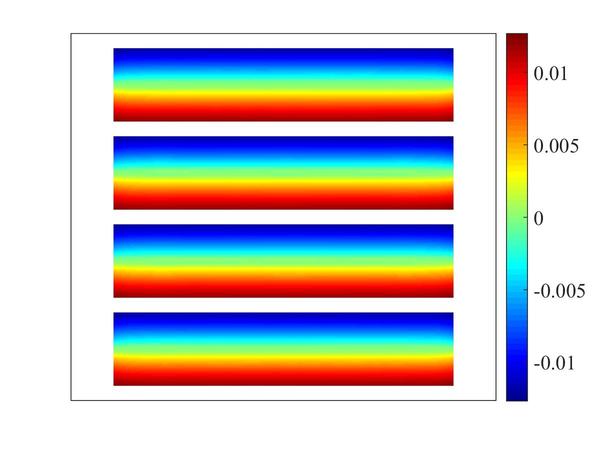} 
		\label{F:E_2D:a}}
\subfigure[Fully coupled stack]{
		\centering
		\includegraphics[scale= 0.2]{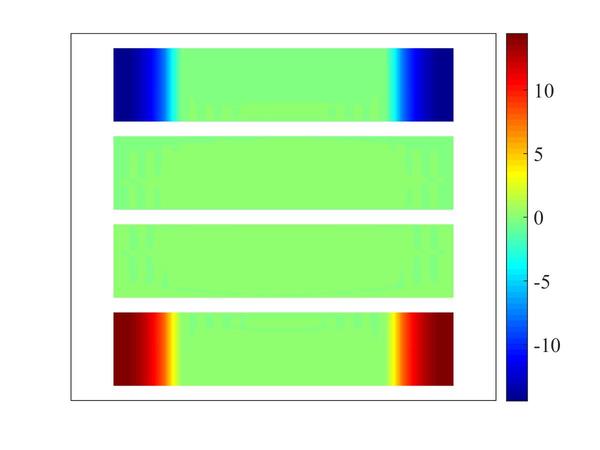} 
		\label{F:E_2D:b}}
\subfigure[Partially coupled stack]{
		\centering
		\includegraphics[scale= 0.2]{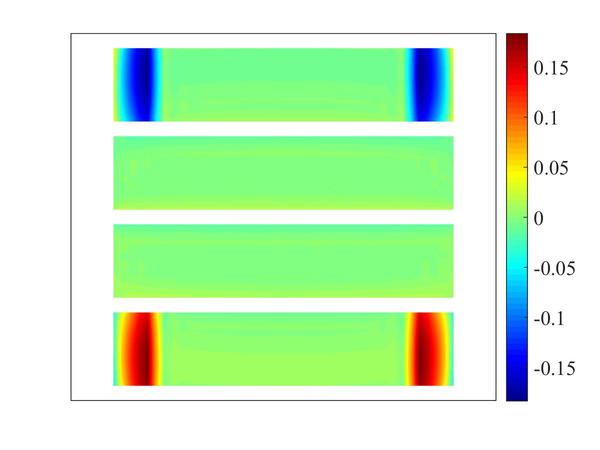} 
		\label{F:E_2D:c}}  
\caption	{ 
Electrical field distribution of uncouple stack, fully coupled stack and partially coupled stack under parallel magnetic field at $B_m=0.02$ T, $f=72$ Hz when $t/T=1.25$, respectively.
The thickness of tape is scaled over 100 times to draw the distribution.  
$(a)$ Uncoupled stack. 
$(c)$ Fully coupled stack.
$(e)$ Partially coupled stack.  
}
\label{F:E_2D}
\end{figure}

\begin{figure}
\centering	
\subfigure[$ t/T = 0.75$]{
	\centering
	\includegraphics[scale= 0.22]{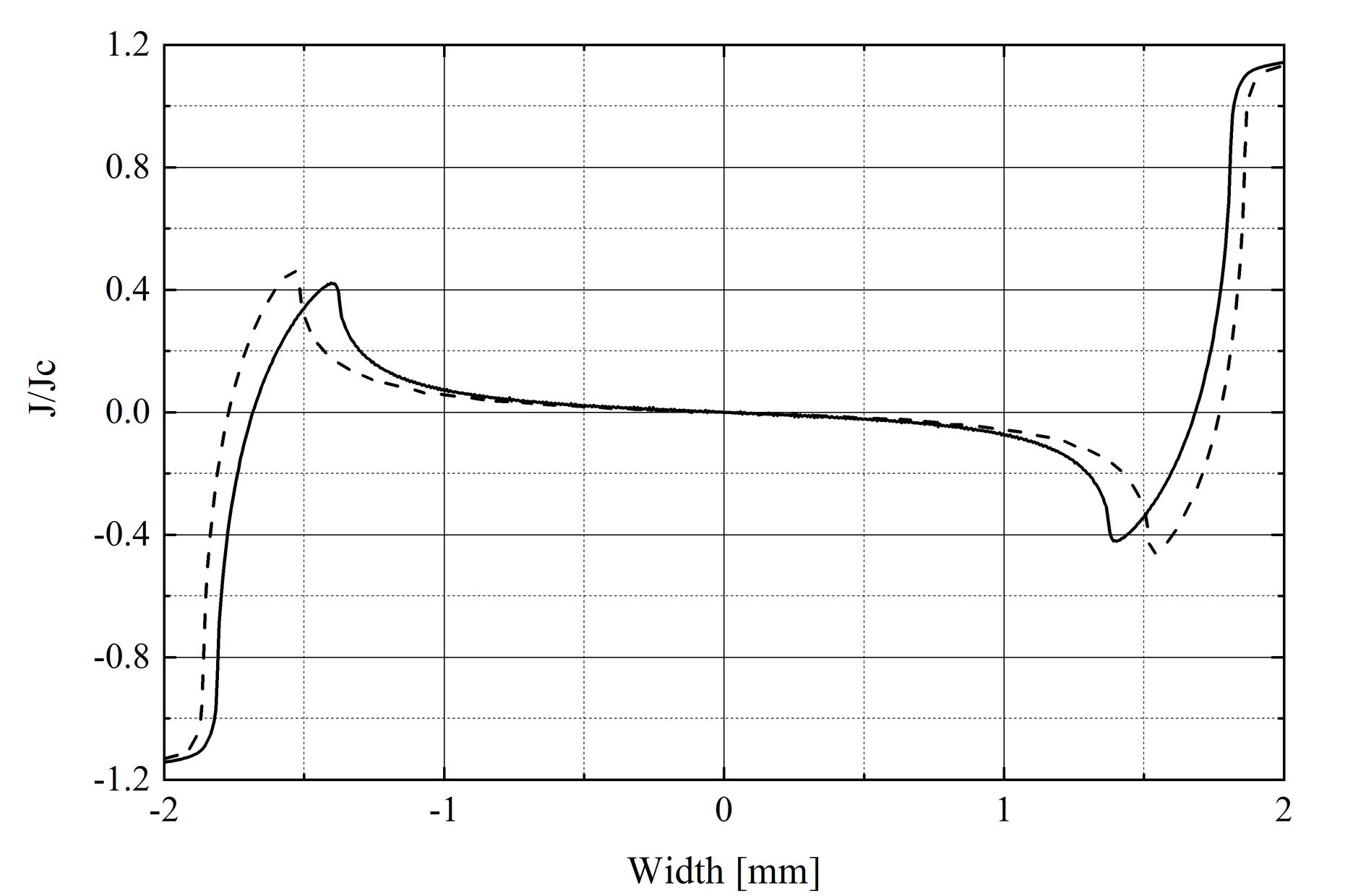} 
	\label{F:E_1C:a}}
\quad
\subfigure[$ t/T = 1$]{
	\centering
	\includegraphics[scale= 0.22]{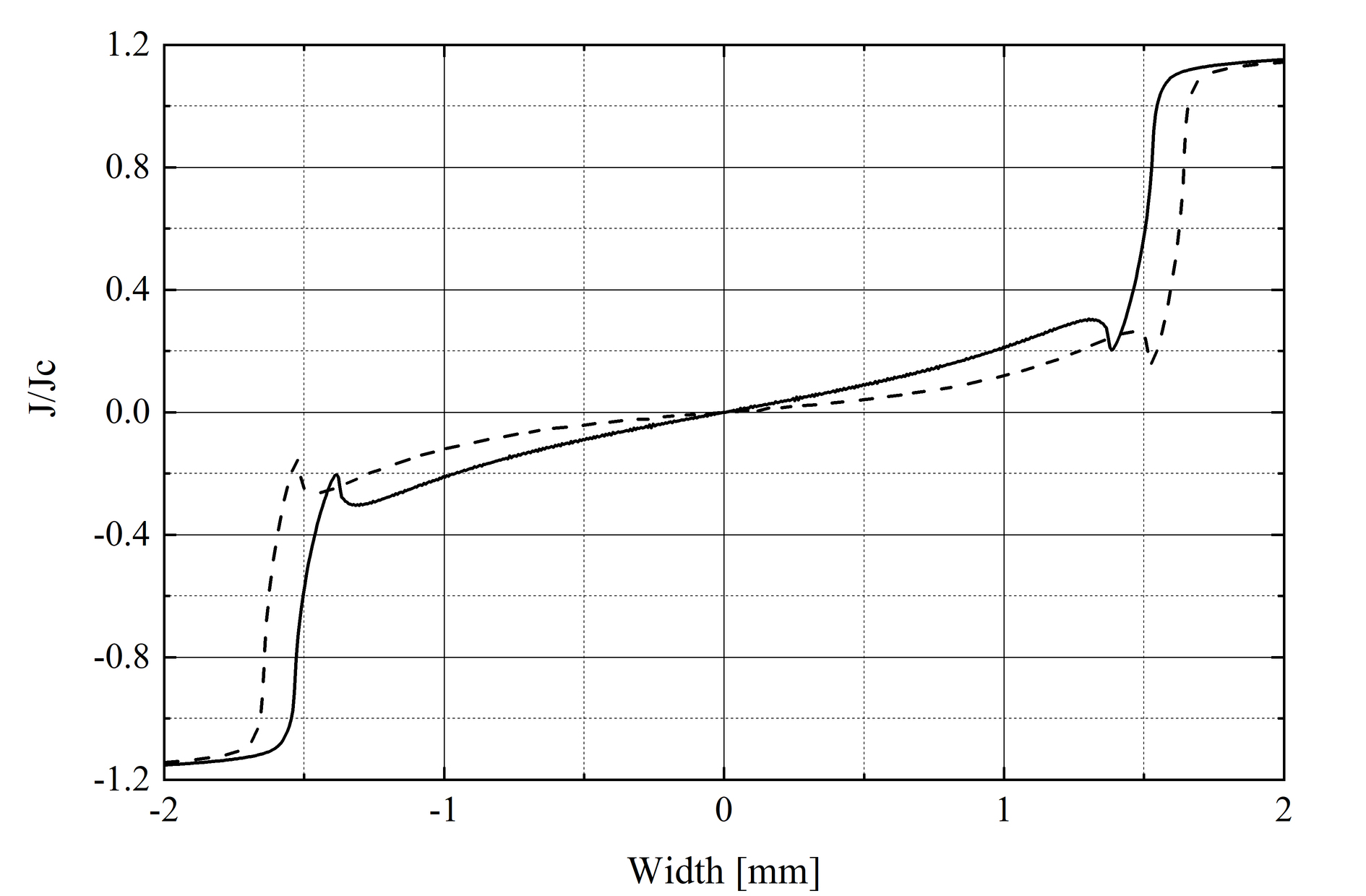} 
	\label{F:E_1C:b}}
\quad
\subfigure[$ t/T = 1.125$]{
	\centering
	\includegraphics[scale= 0.22]{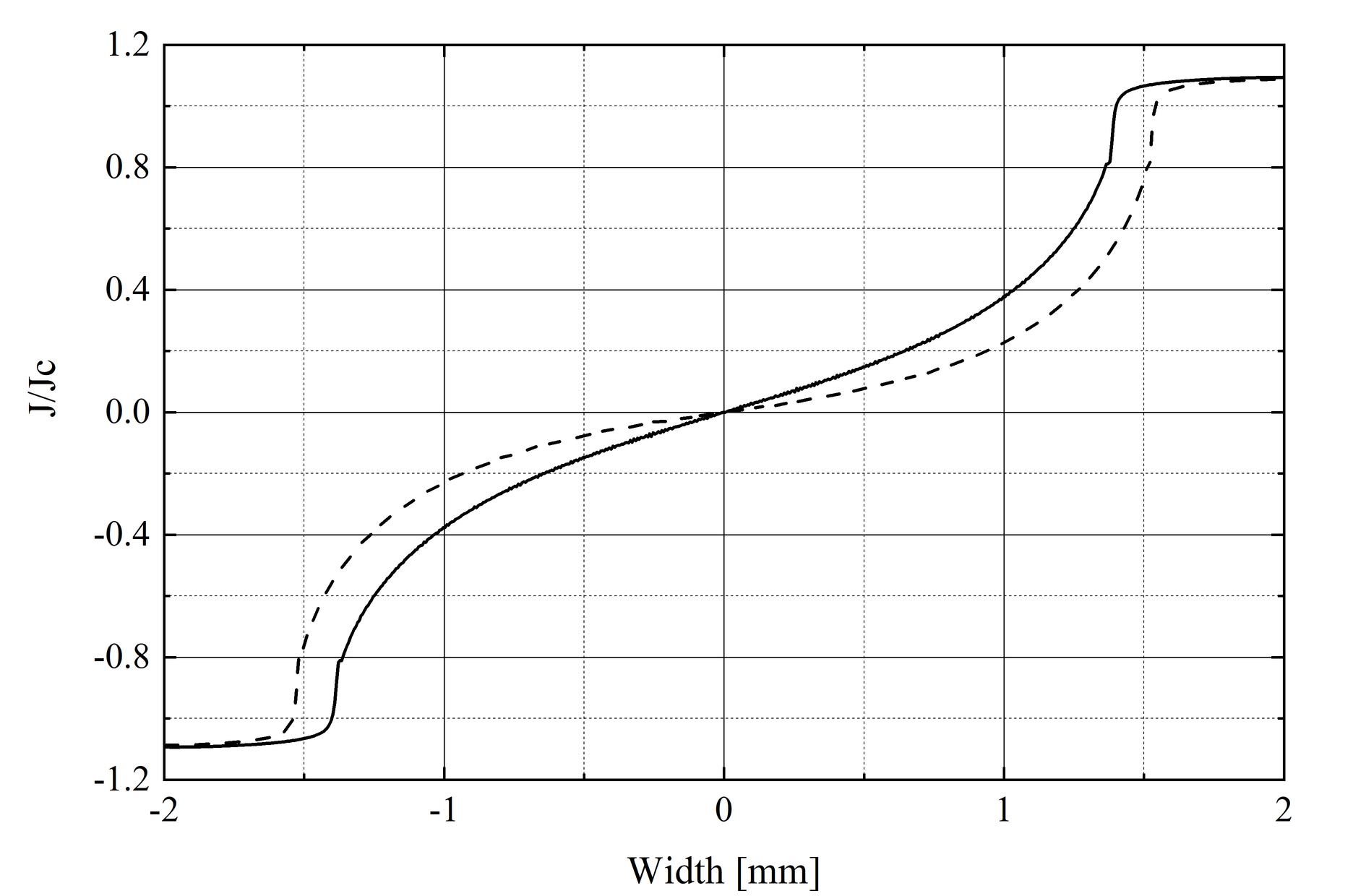} 
	\label{F:E_1C:c}}
\quad
\subfigure[$ t/T = 1.25$]{
	\centering
	\includegraphics[scale= 0.22]{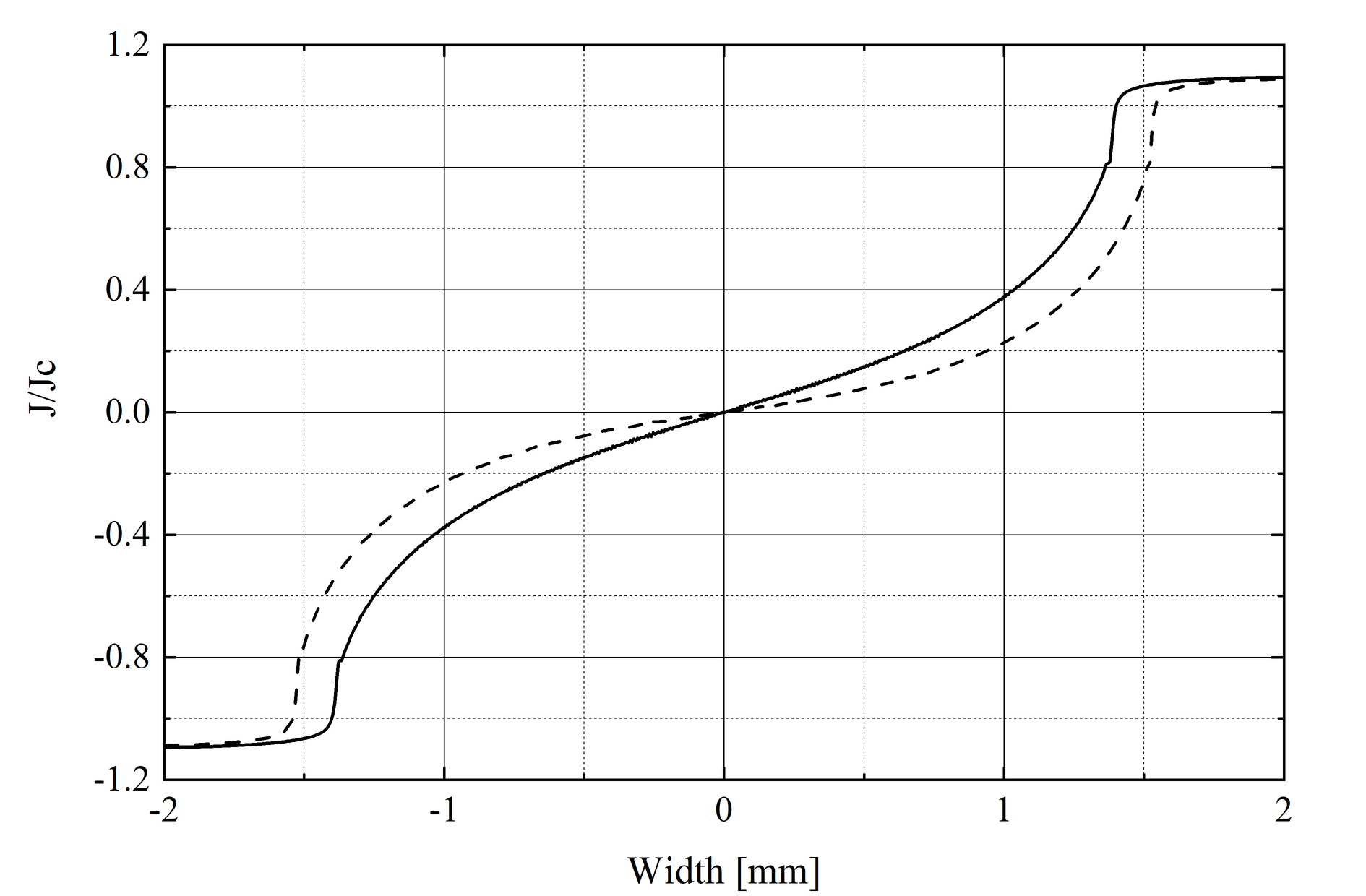} 
	\label{F:E_1C:d}}
\caption	{ 
Current profiles of superconducting stack under perpendicular magnetic field at $B_m = 0.02 $ T,  $f=72$ Hz when $ t/T = 0.75, 1, 1.125$ and $1.25$, respectively. 
The solid lines are the profiles of fist tape and the dash lines are the profiles of second tape. 	
}
\label{F:Cur_1D}
\end{figure}

\begin{figure}
\centering	
\subfigure[$ t/T = 0.75$]{
		\centering
		\includegraphics[scale= 0.22]{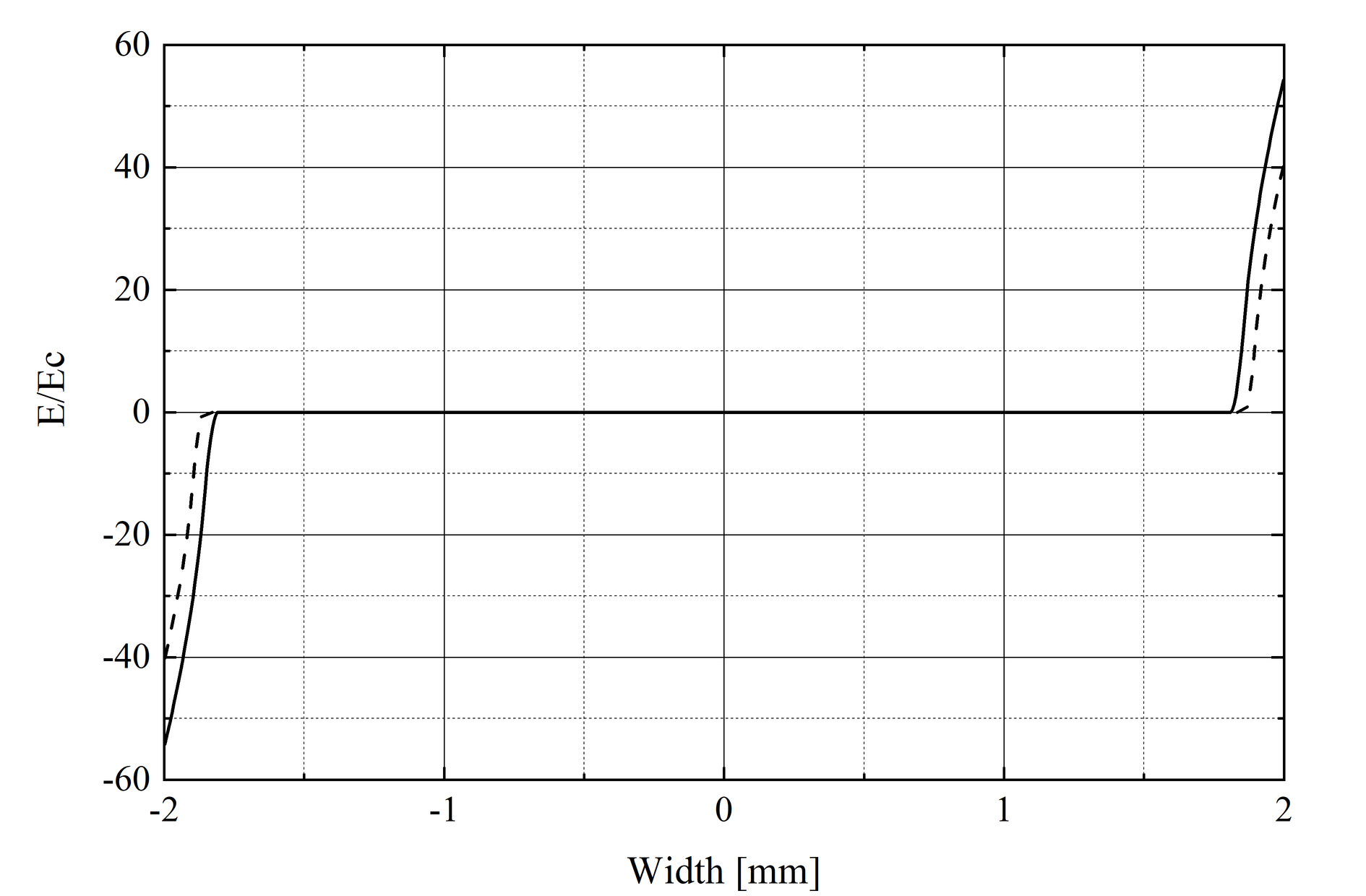} 
		\label{F:E_1D:a}}
\quad
\subfigure[$ t/T = 1$]{
		\centering
		\includegraphics[scale= 0.22]{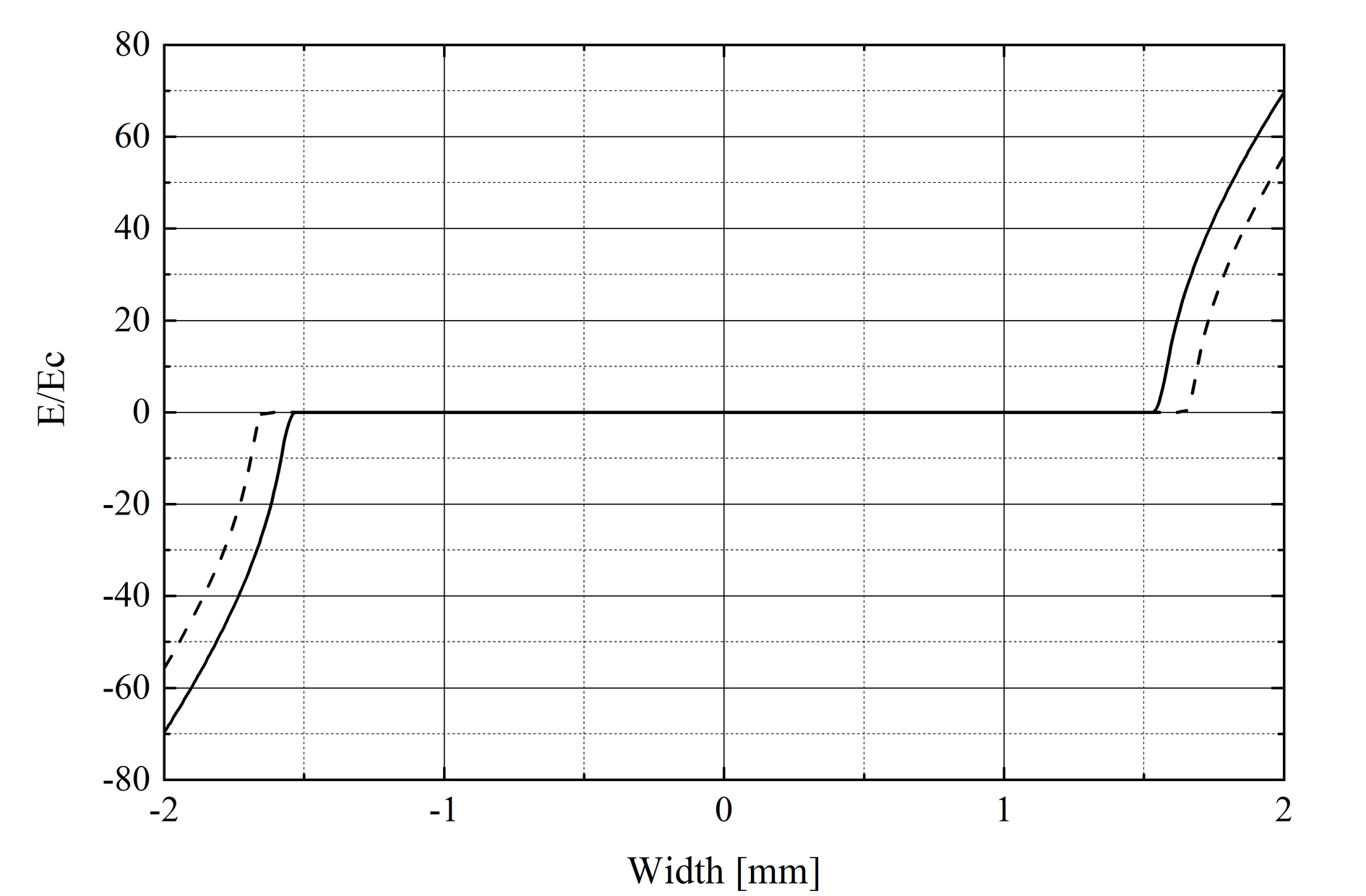} 
		\label{F:E_1D:b}}
\quad
\subfigure[$ t/T = 1.125$]{
		\centering
		\includegraphics[scale= 0.22]{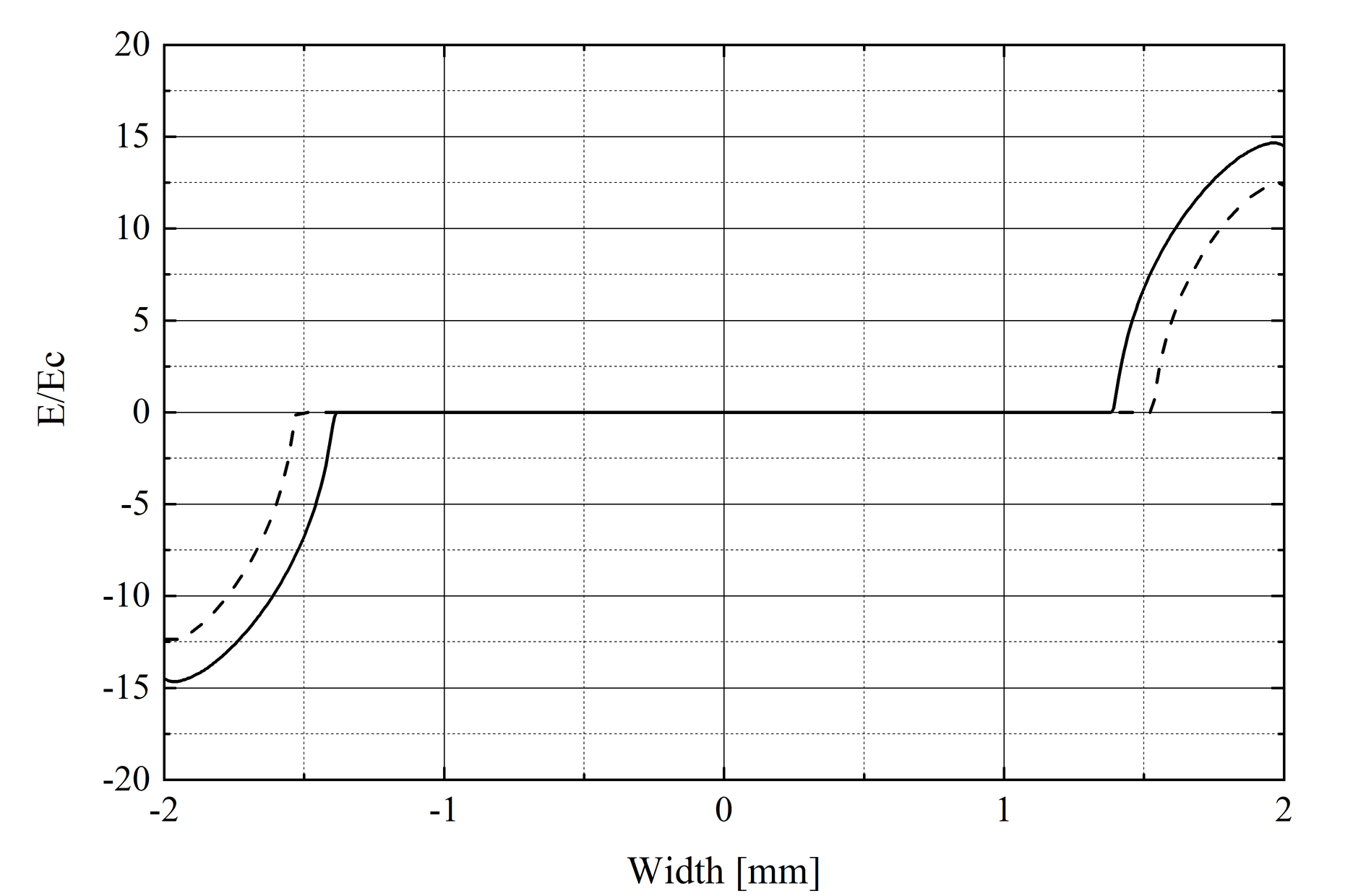} 
		\label{F:E_1D:c}}
\quad
\subfigure[$ t/T = 1.25$]{
	\centering
	\includegraphics[scale= 0.22]{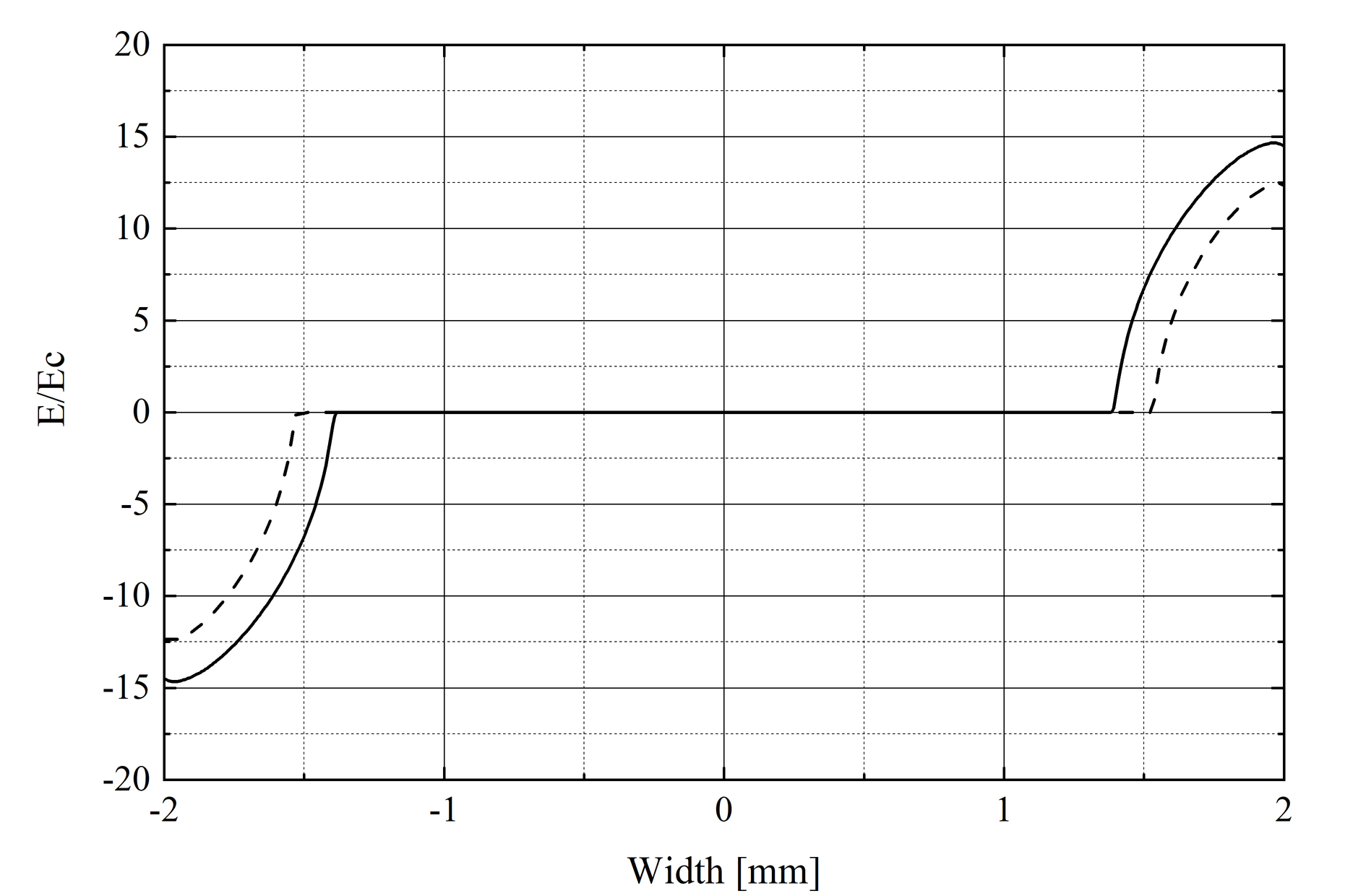} 
	\label{F:E_1D:d}}
\caption	{ 
Electrical field profiles of superconducting stack under perpendicular magnetic field at $B_m = 0.02 $ T,  $f=72$ Hz when $ t/T = 0.75, 1, 1.125$ and $1.25$, respectively. 
The solid lines are the profiles of fist tape and the dash lines are the profiles of second tape. 
}
\label{F:E_1D}
\end{figure}

To show the great difference \R{between the different} kinds of stacks, \R{Figures \ref {F:Cur_2D}-\ref{F:E_1D} present the} current and electrical field distribution and profiles under applied magnetic field, respectively.

\Fref{F:Cur_2D} shows the current distribution of these kinds of stack under parallel magnetic field at $0.02$ T, $72$ Hz when $ t/T = 1.5$. For uncoupled stack,  \Fref{F:Cur_2D:a}, the induced current has fully penetrated into the center of each tape, because each tape is independent \R{with} no physical connection. For fully coupled stack, \Fref{F:Cur_2D:b}, the induced current has penetrated in the first tape and a little bit in the second tape. There is no current in the center of stack, which performs like a superconducting bulk. \R{Partially} coupled \R{stacks}, \Fref{F:Cur_2D:c}, \R{perform} similar to fully coupled \R{stacks} but there are some positive and negative currents in the center of the stack.

\Fref{F:E_2D} shows the electrical field distribution of these kinds of stack under parallel magnetic field. For \R{the} uncoupled stack, \Fref{F:E_2D:a}, \R{the} electrical field distribution is almost the same in each tape. \R{The} peak \R{positions} appear at the thickness direction ($y$ direction).
For \R{the} fully coupled stack, \Fref{F:E_2D:b}, the electrical field is almost zero everywhere except \R{at} the edges. For \R{the} partially coupled stack, \Fref{F:E_2D:c}, the electrical field distribution is similar to \R{the} fully coupled stack but the peak \R{positions} appear at nearby edges and the amplitude of electrical field is much smaller than \R{the} fully coupled stack but larger than \R{the} uncoupled stack.

\Fref{F:Cur_1D} and \Fref{F:E_1D} are the current and electrical field profiles under perpendicular magnetic field when $t/T = 0.75, 1, 1.125, 1.5 $ at amplitude $B_m=0.02$ T \R{and} frequency $f=72$ Hz, respectively. The profiles of first (top) and the second (middle) tapes are drawn. When perpendicular magnetic field \R{is} applied on \R{the} stacks, the induced current flows only in \R{the} $z$ direction and hence there is no coupling current between tapes. Therefore, the current and electrical field profiles of these three superconducting stacks are the same. With time changes from $t/T = 0.75$ to $1.25 $, the amplitude of \R{the} magnetic field increases from \R{the} negative extreme value to \R{the} positive extreme value and the induced current and electrical field penetrates gradually from edges to center.


\subsection{Frequency dependence}

\begin{figure}
\centering
\includegraphics[scale= 0.4]{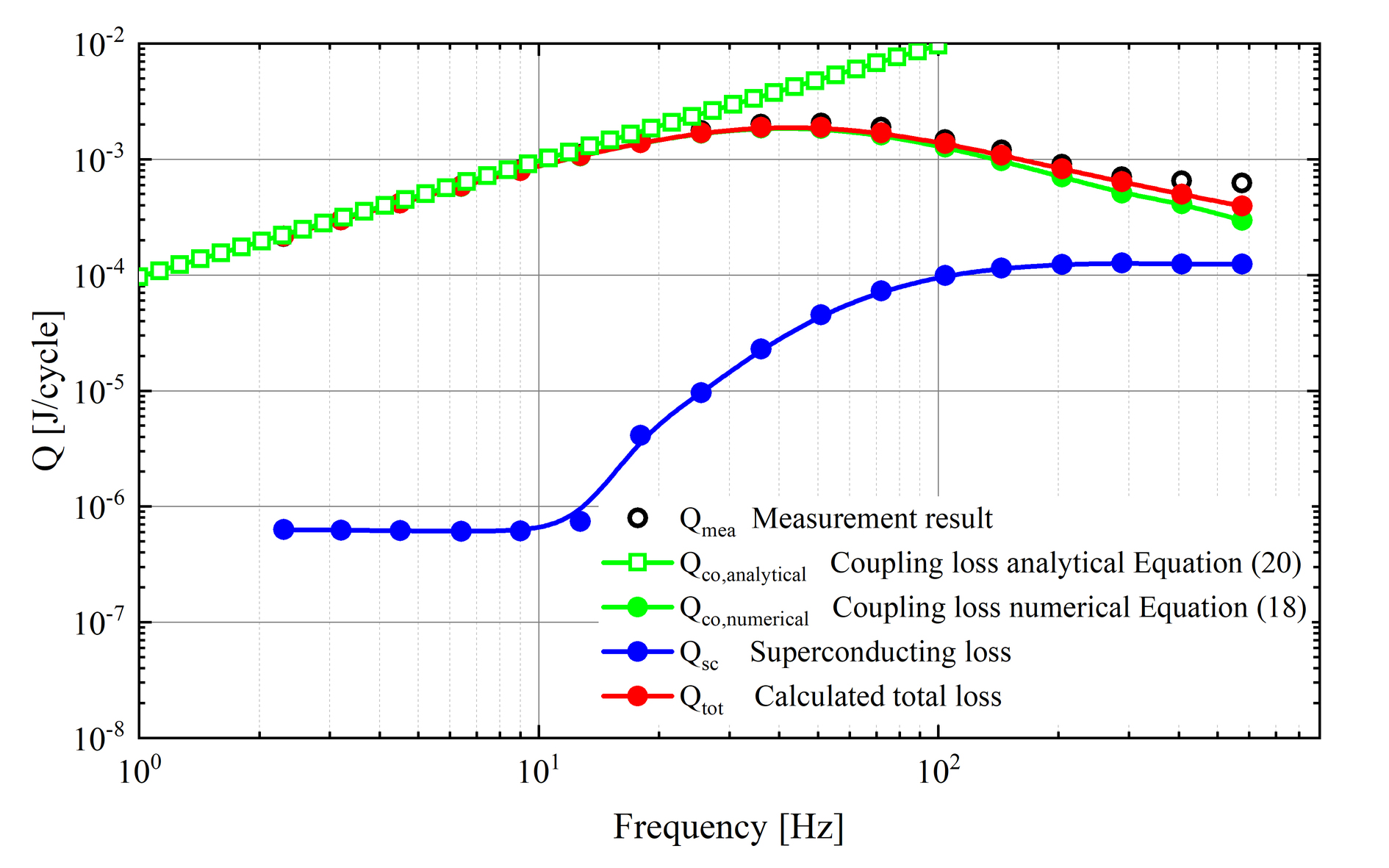} 
\caption	{	
Magnetization loss of \R{the} partially coupled stack under parallel magnetic field at $18.7$ mT with frequency ranging from $2$ Hz to $576$ Hz. 
The black circle line is measured total loss $Q_{mea}$,
and the green open rectangle line is coupling loss $Q_{co,analytical}$ calculated by analytical method \Eref{EA:Q}.
Others are numerical result calculated by \Eref{E:Q}:
the green solid circle line is coupling loss $Q_{co,numerical}$ \R{(or resistor loss $Q_R$ in the text)},
the blue solid circle line is superconducting loss $Q_{sc}$ \R{(or $Q_s$ in the text)}.
and the red solid circle line is calculated total loss $Q_{tot}$.
}
\label{F:frequency}
\end{figure}

\begin{figure}
	\centering
	\includegraphics[scale= 0.4]{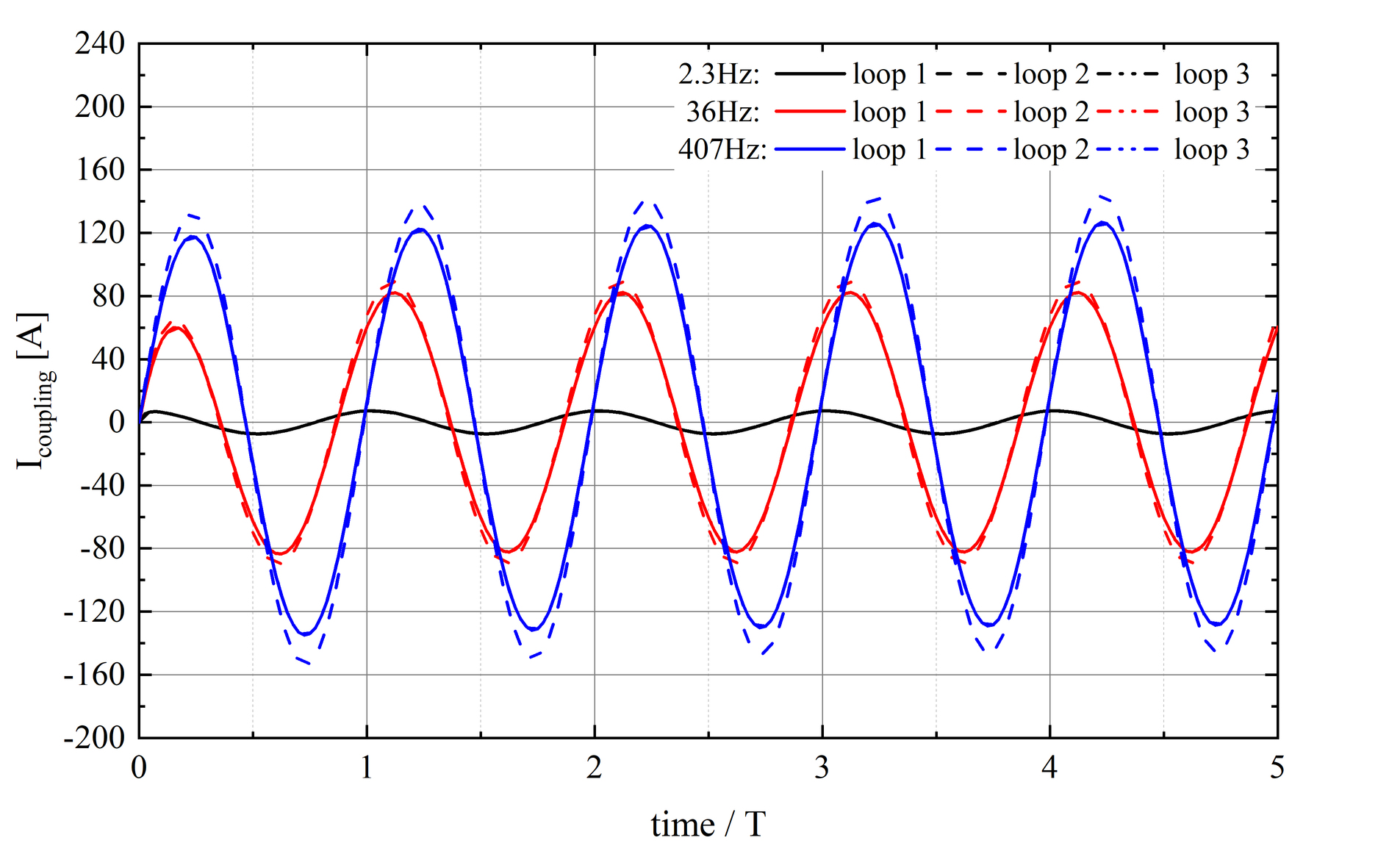} 
	\caption	{	
	Coupling current of partially coupled stack under parallel magnetic field at $18.7$ mT with frequency  $2.3, 36,$ and $407$ Hz.
	The	solid line is the coupling current in loop 1, the dash line is the coupling current in loop 2, and the double dash line is the coupling current in loop 3. Because of the geometry symmetry of the stack, induced current in loop 1 and loop 3 are overlap each other.
	}
	\label{F:current_frequency}
\end{figure}

\Fref{F:frequency} shows the superconducting loss and coupling loss of \R{the} partially coupled stack sample under parallel magnetic field. \R{The} coupling current at frequency $2.3$ Hz,$36$ Hz and $407$ Hz are shown in \Fref{F:current_frequency}. The amplitude of parallel applied magnetic field is $18.7 $ mT and the frequency changes  from $2$ Hz to $576$ Hz. When the frequency is below $10$ Hz, the superconducting loss in \R{the} tapes $Q_{s}$ is almost no frequency independent. When the frequency is above $10$ Hz, $Q_{s}$ increases sharply but \R{it} is still hundreds of times lower than the \R{resistor} loss $Q_{R}$ in \R{the} soldering parts. However, when the frequency is higher than $100$ Hz, $Q_{s}$ increases much slower than before and trends to constant. Meanwhile, $Q_{R}$ increases at \R{the} low frequency region and then decreases at \R{the} high frequency region. The maximum coupling loss appears at around $40$ Hz, \R{where the position of this maximum} is related to the resistance and inductance. 

The frequency dependence is mainly contributed by the \R{resistances} between tapes. The numerical result $Q_{tot}$ is consistent with \R{the} measurement result $Q_{mea}$ with \R{a} maximum relative error below $6 \%$, which happens at high \R{the} frequency region. Furthermore, an analytical formulation of coupling loss for low frequency is deduced in \Eref{EA:power}, which \R{agrees} with numerical calculations and measurements at low frequency. However, when the frequency is higher than $300$ Hz, the relative error is larger. The reason could be the eddy currents closing entirely within the soldering parts, which the model does not take into account. Another source of error could be inaccuracies in the superconducting loss calculation due to the constant $J_c$, which causes an increase in the superconducting loss.

\subsection{Resistance dependence}

\begin{figure}
\centering
\includegraphics[scale= 0.4]{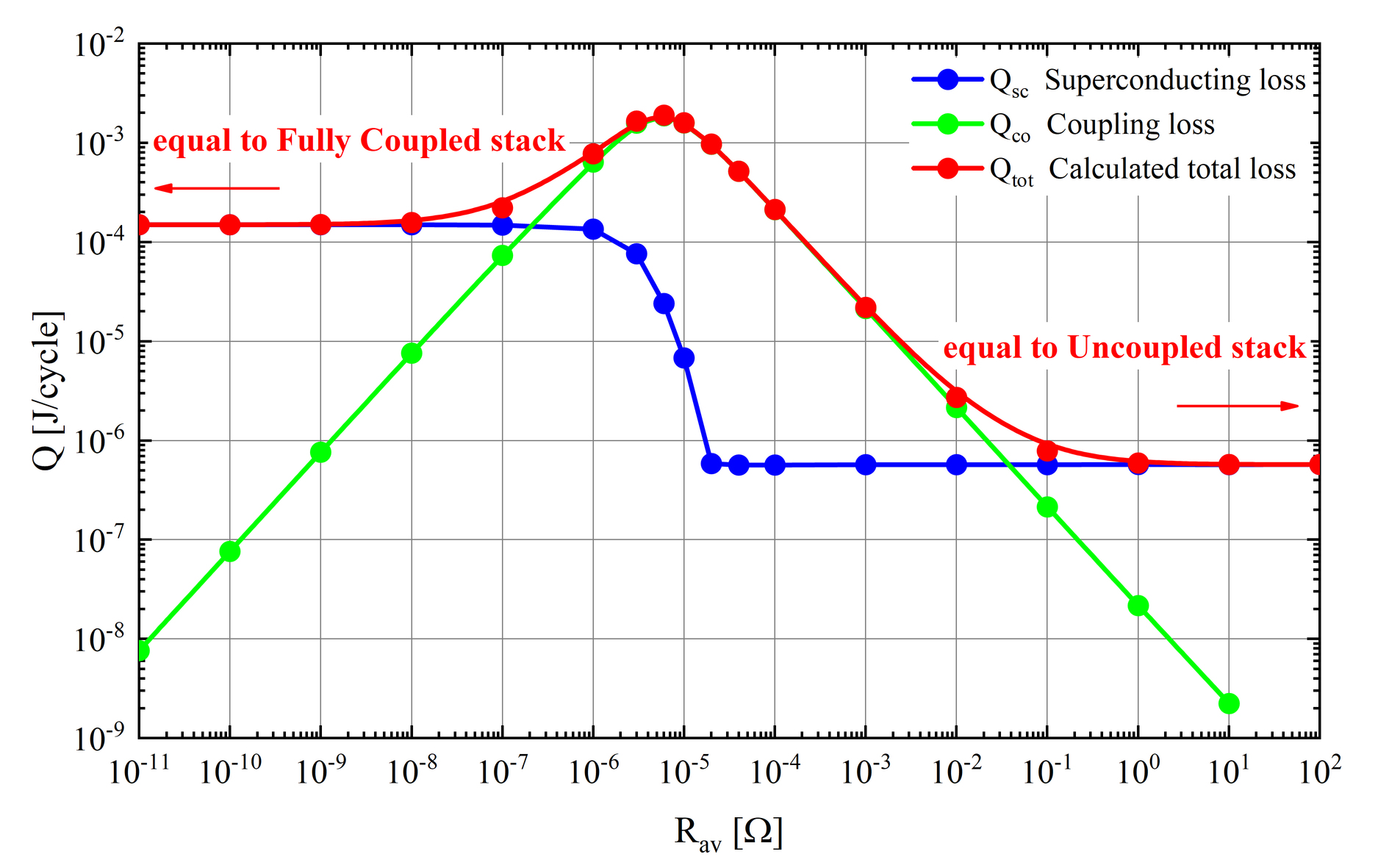} 	
\caption	{
Magnetization loss of partially coupled stack under parallel magnetic field at $18.7$ mT $72$ Hz with average resistance ranging from $ 1\times 10^{-9}$ $ \Omega $ to $ 1\times 10^{2}$ $ \Omega $. 
The blue line is superconducting loss $Q_{sc}$, the green line is coupling loss $Q_{co}$, and the red line is the total loss $Q_{tot}$.
}
\label{F:resistance}
\end{figure}


\begin{figure}
	\centering
	\includegraphics[scale= 0.4]{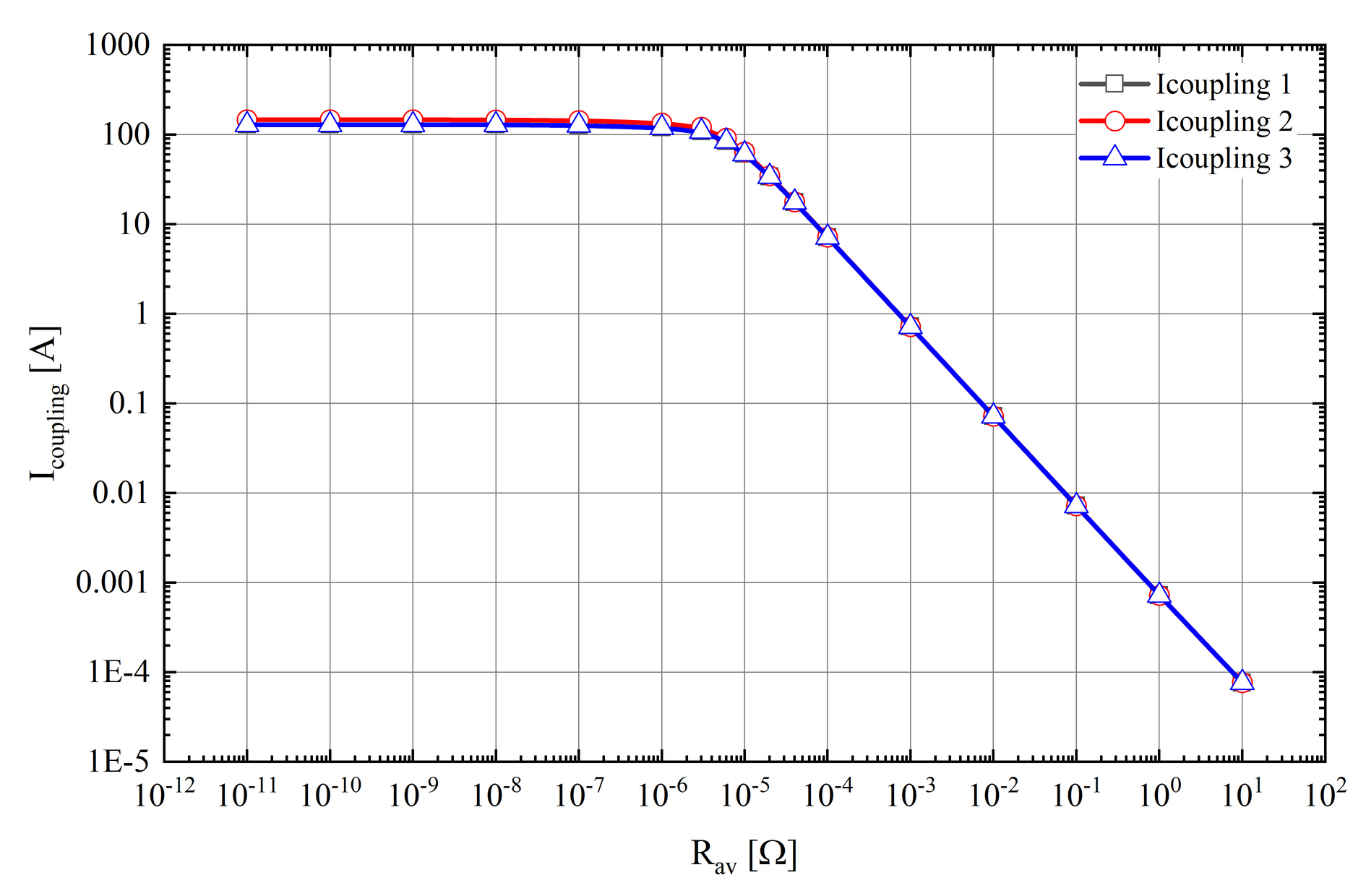} 	
	\caption	{
		Coupling current (the amplitude per cycle) of partially coupled stack under $18.7$ mT, $72$ Hz parallel magnetic field with average resistance ranging from $ 1\times 10^{-9}$ $ \Omega $ to $ 1\times 10^{2}$ $ \Omega $. 
		The black line is the coupling current in loop 1. 
		The red line is the coupling current in loop 2. 
		The blue line is the coupling current in loop 3. 	
	}
	\label{F:current_resistance2}
\end{figure}

\Fref{F:resistance} shows the superconducting loss and coupling loss of the partially coupled stack sample under parallel magnetic field with amplitude $18.7 $ mT and frequency $72$ Hz. The resistance changes from $1 \times 10^{-10}$ $  \Omega$ to $1 \times 10^{2}$ $  \Omega$. \Fref{F:current_resistance2} shows the coupling current of the partially coupled stack. Because \R{of} the \R{geometrical} symmetry of the stack, coupling currents in loop 1 and 3 are overlapped \R{with} each other. \R{However,} the coupling current in loop 2 is higher than \R{the} others at \R{the} low resistance region. 

With low resistances ($R_{av} < 1 \times 10^{-6}$ $\Omega$), the net current flowing between tapes is almost constant\R{,} which is entirely due to inductive effects\R{. The} value of the resistance is not playing any role, \R{and hence the resistor loss $Q_{R}$} decreases with the resistance $R_{av}$. The superconducting loss in \R{the} tapes $Q_{s}$ is almost constant, which is about $2 \times 10^{-4}$ J/cycle, \R{while the resistor} loss in \R{the} soldering parts $Q_{R}$ is several orders of magnitude smaller than $Q_{s}$. Since \R{the} coupling current is fixed, see \Fref{F:current_resistance2}, the Joule loss in the resistance $R_{av}I_{coupling}^2$ increases linearly with the resistance. Intuitively, the case of $R_{av} = 0 $ $ \Omega$ results in no \R{resistor} loss $Q_{co}=0$ J/cycle, since it corresponds to perfect shielding. \R{Therefore}, the total loss of the partially coupled stack $Q_{tot}$ is almost the same as \R{the} fully coupled stack.

When the resistance is around $7 \times 10^{-6}$ $ \Omega$, $Q_{R}$  reaches the peak which is about $2 \times 10^{-3}$ J/cycle. However, when the resistance is around  $1 \times 10^{-6}$ $  \Omega$ $ \sim 2 \times 10^{-5}$ $ \Omega $, \R{the} coupling current decreases to zero sharply and $Q_{sc}$ drops over $200 $ times. \R{Since} the coupling current is now dominated by the resistance\R{, the coupling current decreases with  $R_{av}$.} This decrease is linear because the coupling current is so small that the applied magnetic field freely flows between tapes, causing an induced voltage independent on the coupling current and $R_{av}$. At last, coupling current \R{cannot} flow anywhere, and \R{hence} the total loss $Q_{tot}$ is \R{corresponds to that} of the uncoupled stack.

\subsection{Length dependence}

\begin{figure}
\centering
\includegraphics[scale= 0.4]{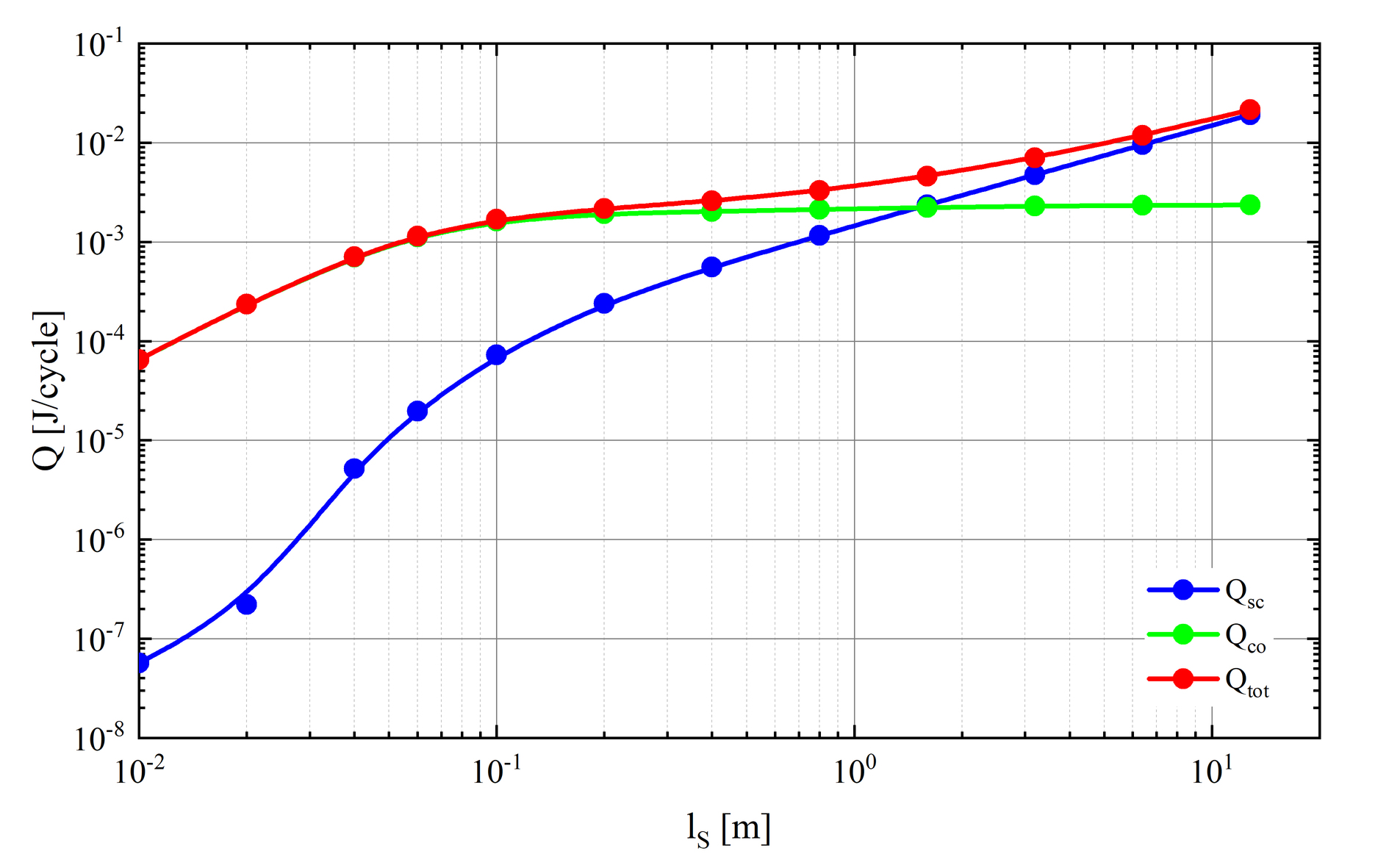} 
\caption	{
Magnetization loss of \R{the} partially coupled stack under $18.7$ mT, $72$ Hz parallel magnetic field with length ranging from $10$ mm to $12800$ mm.
The black line is magnetization loss $Q_{sc}$, 
the red line is coupling loss $Q_{co}$. 
And the blue line is total loss $Q_{tot}$.
The loss unite is J/cycle.
}
\label{F:length}
\end{figure}

\begin{figure}
	\centering
	\includegraphics[scale= 0.4]{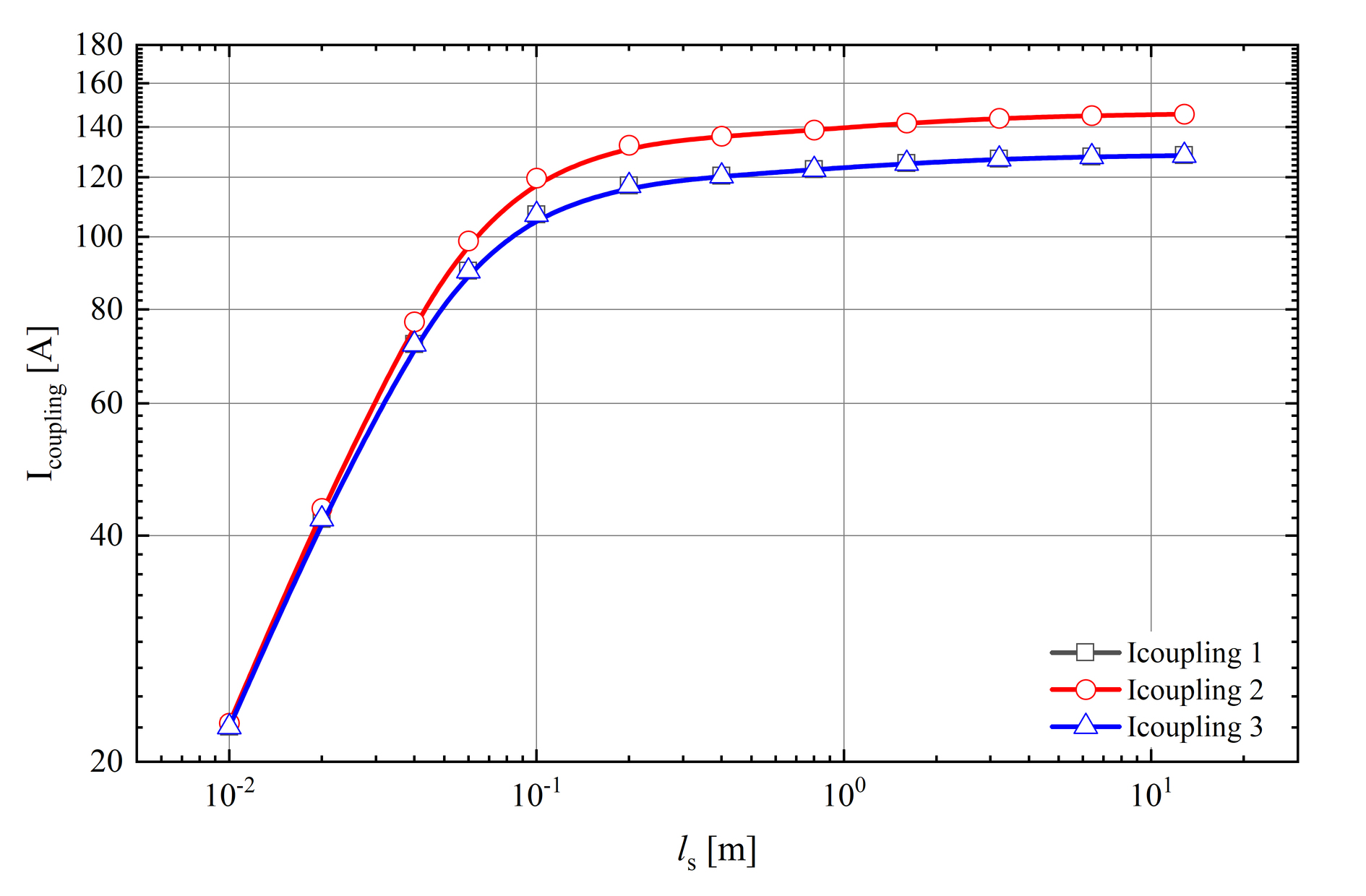} 
	\caption	{
		Coupling current in loops with length ranging from $10$ mm to $12800$ mm.
	}
	\label{F:length_Cur}
\end{figure}

\Fref{F:length} shows the magnetization loss of \R{the} partially coupled stack and \Fref{F:length_Cur} is the coupling current with length increasing from $10$ mm to $12800$ mm. When the length is smaller than $100$ mm, the superconducting loss in \R{the} tapes $Q_{s}$ is about one thousand times smaller than $Q_{R}$. \R{Since} the resistance is constant and the coupling current increases linearly \R{with} the length, $Q_{R}$ \R{increases following} slope $2$ \R{in log-log scale}. When the length \R{is} larger than $100$ mm\R{, the} induced current in \R{the} superconducting tape trends to constant, \R{and hence} $Q_{R}$ trends to constant \R{but} $Q_{s}$ increases in proportion to \R{the} length. \R{When the length further increases}, the normalized total loss $Q_{tot}$ \R{is} mainly contributed by $Q_{s}$.

\section{Conclusion}\label{sec:conclusion}

A numerical 2D model is developed by \R{the} minimum electromagnetic entropy production \R{(MEMEP)} method to compute the coupling loss of partially coupled \R{stacks}. The presented MEMEP model shows the capability to take the resistance between tapes into account for coupling loss calculation with a high number of mesh \R{elements}. \R{In addition,} this model \R{provides} a remarkable good agreement with \R{measurements}.

This model is later used to systematically study the amplitude dependence, frequency dependence, resistance dependence, and length dependence of coupling loss. When a sinusoidal perpendicular magnetic field \R{is} applied on \R{the} stack samples, these superconducting stacks perform \R{the} same features and no obvious frequency dependence \R{is} observed. When a sinusoidal parallel magnetic field \R{is} applied, the total loss of partially \R{coupled} stack is much higher than \R{the} uncoupled stack. This is because there are resistances at \R{the} soldering parts that induce coupling loss.
Obvious frequency dependence, resistance dependence and length dependence are detected in partially coupled stack too. With \R{increasing} frequency (or resistance), the coupling loss of \R{the} partially coupled stack increases at \R{the} beginning and then decreases. With the length of the sample increasing, the superconducting loss increases sharply at \R{the} beginning and then increases in proportion to \R{the} length, but the coupling loss in \R{the} soldered parts trends to constant.  

At last, the 2D model presented here is useful for determining not just the value of predict the coupling loss but also other electromagnetic behavior in samples that are connected only at the ends, such as certain stacks of tapes, or coils with parallel tapes as conductor. \R{This fast and accurate 2D model enables to study the effect of coupling at the current terminals of superconducting coils in complex systems, such as superconducting motors, where a full 3D model taking these efects into account is unfeasible due to its huge computing complexity.}

\section*{Acknowledgements}

This paper is supported by European Commission Project Horizon 2020, No.723119 \R{(ASuMED)}.


\section*{References}
\bibliographystyle{unsrt}






\appendix
\section{Coupling loss analytical modeling for low frequency} \label{sec:app_low}

\begin{figure}
	\centering
	\includegraphics[scale= 0.4]{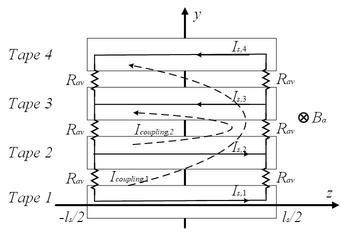} 	
	\caption	{
		Sketch of partially coupled stack for analytical modeling. (cross section in y-z platform )	
	}
	\label{FA:sketch}
\end{figure}

In this section, coupling loss of partially coupled stack at low frequency is deduced.
There are four tapes in stack.

When a parallel magnetic field is applied on the partially coupled stack, the coupling current is induced which obey Ampere' Law.
\begin{eqnarray}\label{key0}
\oint d\vec{S} \cdot \vec{B}_a = L\dot{I}+ \dot{B}_a \cdot d \cdot l_s 	
\end{eqnarray}
where $d$ is the gap between tapes. 
$l_s $ is the length of tape. 
$L$ is the inductance of the superconductor loop which is neglect-able at low frequency.
Based on Faraday' Law,
\begin{eqnarray}	\label{RI0}
R_{av}I = - \frac{1}{2} \cdot d \cdot l_s  \cdot \dot{B}_a -\vec{E}_s \cdot l_s 
\end{eqnarray}
Because $I << I_c$, the electromotive $E_s$ on superconducting tapes approximate zero. 
So,
\begin{eqnarray}  \label{RI1}
R_{av}I = - \frac{1}{2} \cdot d \cdot l_s  \cdot \dot{B}_a
\end{eqnarray}
where $B_a$ is parallel applied magnetic field, $B_a= B_m \sin \omega t$. 
And $\dot{B}_a=\partial B_a / \partial t= \omega B_m \cos \omega t$, $B_m$ is the amplitude of the applied parallel magnetic field.

As shown in \Fref{FA:sketch}, 
because of the geometry symmetry of the partially coupled stack, tape No. 1 and No. 4 form a big current loop and tape No. 2 and No. 3 form another current loop.
So, induced currents in tapes are written as
\begin{eqnarray}
I_{s,1} & = -\frac{l_s }{2R_1}(\dot{B}_a \cdot d_1 )  \\	
I_{s,2} & = -\frac{l_s }{2R_2}(\dot{B}_a \cdot d_2 ) 
\end{eqnarray}
And induced current in tape No.3 and No. 4 are written as $I_{s,4}=-I_{s,1}$ and $I_{s,3}=-I_{s,2}$, respectively.

Because $R_1=3R_{av}, R_2=R_{av}, d_1=3d$, and $d_2=d$, power loss in resistance is deduced,	
\begin{eqnarray}
P(t) & = 2(R_1-R_2)I_{s,1}^2+2R_2 (I_{s,1}+I_{s,2})^2 \nonumber\\
&=\frac{3}{2R_{av}}(l_s \cdot d \cdot \omega {B}_m \cos \omega t)^2 
\end{eqnarray}

And average power loss is
\begin{eqnarray}	
<P> = \frac{3\omega ^2}{4R_{av}}(l_s \cdot d \cdot  {B}_m)^2 
\end{eqnarray}
\begin{eqnarray}	
<Q> =<P>/f= \frac{3 \pi \omega}{2R_{av}}(l_s \cdot d \cdot {B}_m)^2 
\end{eqnarray}


\section{Coupling loss analytical modeling for any frequency} \label{sec:app_high}
In this section, coupling loss of partially coupled stack at high frequency is deduced by analytical method.
To simplify the issue,  only two superconducting tapes in closed loop are considered.

At high frequency, the inductance is not negligible anymore.
\begin{eqnarray}
L= d \cdot l_s \cdot  \mu _0 /W
\end{eqnarray}
where, $W$ is the width of tape. 
$d$ is the gap between tapes. $l_s $ is the length of tape. 

\Eref{RI1} is revised to
\begin{eqnarray}	 \label{RI2}
R_{av}I+ \frac{1}{2} L \dot{I} = -\frac{1}{2} \cdot d \cdot l_s  \cdot \dot{B}_a	
\end{eqnarray}
where, $R_{av}$ is the average resistance in closed loop. And for any current can be written in Fourier decomposition form,
\begin{eqnarray} \label{I}
I  & = I_{co} \cos \omega t  + I_{si} \sin \omega t   \\
\dot{I}   & = -\omega I_{co} \sin \omega t + \omega I_{si} \cos \omega t  \nonumber
\end{eqnarray}
where $I_{co}$ and $I_{si}$ are two components of the induced current.

Take \Eref{I} into \Eref{RI2} to solve the $I_{si}$ and $I_{co}$.
\begin{eqnarray}
\begin{cases} 
I_{si}-\frac{\omega L}{2R_{av}}\cdot  I_{co} =0  \\
I_{co}+\frac{\omega L}{2R_{av}} \cdot  I_{si}+\frac{1}{2R_{av}} \cdot d \cdot l_s \cdot \omega B_m=0
\end{cases}
\end{eqnarray}

So, we can deduce
\begin{eqnarray} \label{Ico}
I_{co} = -\frac{1}{2R_{av}}\cdot d \cdot l_s   \cdot \omega B_m \cdot \frac{1}{1+(\frac{L \omega}{2R_{av}})^2} 
\end{eqnarray}
\begin{eqnarray} \label{Isi}
I_{si} = -\frac{1}{2R_{av}} \cdot \frac{\omega L}{2R_{av}} \cdot d \cdot l_s  \cdot \omega B_m \cdot \frac{1}{1+(\frac{L \omega}{2R_{av}})^2} 
\end{eqnarray}

Then, the coupling power is
\begin{eqnarray}
P(t)  
& = 2R_{av}I^2 \nonumber\\
& = 2R_{av} \cdot (I_{co} \cos \omega t +I_{si} \sin \omega t )^2   \nonumber\\
& = \frac{ (\cos \omega t + \frac{L \omega}{2R_{av}} \cdot \sin \omega t )^2}{2R_{av}} \cdot \left(\frac{d \cdot l_s  \cdot \omega \cdot B_m}{1+(\frac{L \omega }{2R_{av}})^2}  \right)^2 
\end{eqnarray}

So, average coupling loss is
\begin{eqnarray}
<P>=\frac{1}{4R_{av}} \cdot (d \cdot l_s  \cdot \omega \cdot B_m)^2 \cdot \frac{1}{1+(\frac{L \omega }{2R_{av}})^2} 
\end{eqnarray}
\begin{eqnarray}
<Q> &= <P>/ f  \nonumber \\
&=	\frac{\pi \omega}{2R_{av}} \cdot (d \cdot l_s  \cdot B_m)^2 \cdot  \frac{1}{1+(\frac{L \omega }{2R_{av}})^2} 
\end{eqnarray}

{\bf $\blacksquare$ Amplitude dependency}

Because,
\begin{eqnarray}	 \label{E:dQ/dB}
\frac{d<Q>}{d B_m}
&=	\frac{\pi \omega \cdot (d \cdot l_s )^ 2}{R_{av}}  \cdot  \frac{B_m}{1+(\frac{L \omega }{2R_{av}})^2}>0
\end{eqnarray}
The coupling loss is always increasing linearly with amplitude at a specific rate which is related to $\omega$ and $R_{av}$.

{\bf $\blacksquare$ Frequency dependency}

When $ \omega =2R_{av}/L$,  ${d<Q>}/{d \omega}=0$. 	
\begin{eqnarray}	 \label{E:dQ/dw}
\frac{d<Q>}{d \omega}
&=	\frac{\pi \cdot (d \cdot l_s  \cdot B_m)^2}{2R_{av}}  \cdot \frac{1-(\frac{L\omega}{2R_{av}})^2}{[1+(\frac{L \omega }{2R_{av}})^2]^2} \nonumber \\
&=0	
\end{eqnarray}
 
The peak value of coupling loss is
\begin{eqnarray}  \label{E:Qf}
<Q_{co}>_f &= \frac{\pi}{2L} \cdot (d \cdot l_s  \cdot B_m)^2
\end{eqnarray}
The peak value of $<Q_{co}>_f$ is only related to the geometry of the stack and the amplitude of applied magnetic field  but is not related to $R_{av}$. 
So, once the geometry is fixed, the peak value of frequency dependency is constant.

{\bf $\blacksquare$ Resistance dependency}

When $R_{av}=L \omega /2$, $d<Q>/d R_{av}=0$.
\begin{eqnarray}	\label{E:dQ/dR}
\frac{d<Q>}{d R_{av}}
&=	\frac{\pi \omega \cdot (d \cdot l_s  \cdot B_m)^2}{2}  \cdot \frac{(\frac{L \omega}{2})^2-R_{av}^2}{[R_{av}^2+(\frac{L \omega }{2})^2]^2} \nonumber \\
&=0	
\end{eqnarray} 

So, the peak value of coupling loss is
\begin{eqnarray}	 \label{E:QR}
<Q_{co}>_R=	\frac{\pi}{L} \cdot (d \cdot l_s  \cdot B_m)^2 
\end{eqnarray} 
The peak value of $<Q_{co}>_R$ is only related to the geometry of the stack and the amplitude of applied magnetic field but is not related to $\omega$.

{\bf $\blacksquare$ Gap ($d$) dependency or length ($l_s $) dependency}
\begin{eqnarray} \label{E:dQ/dd}	
\frac{d<Q>}{d d}=	\frac{\pi \omega  \cdot (l_s  \cdot B_m)^2 }{2R_{av}}  \cdot \frac{2d}{[1+(\frac{L \omega }{2R_{av}})^2 ]^2} \ge 0	
\end{eqnarray}
\begin{eqnarray}\label{E:dQ/dls}	
\frac{d<Q>}{d l_s}=	\frac{\pi \omega  \cdot (d  \cdot B_m)^2}{2R_{av}}  \cdot \frac{2l_s}{[1+(\frac{L \omega }{2R_{av}})^2]^2} \ge 0	
\end{eqnarray}

so, $<Q>$  is always monotone increasing with gap and length.


\end{document}